\documentclass[twocolumn, twocolappendix]{aastex63}

\shorttitle{RAGN detected in a protostructure at z$\sim$3.3}
\shortauthors{Shen et al.}

\begin{document}

\title{Implications of the Environments of Radio-detected AGN in a Complex Protostructure at z$\sim$3.3}

\correspondingauthor{Lu Shen, Guilin Liu, Wenjuan Fang}
\email{lushen@ustc.edu.cn, glliu@ustc.edu.cn, wjfang@ustc.edu.cn,}

\author[0000-0001-9495-7759]{Lu Shen}
\affil{CAS Key Laboratory for Research in Galaxies and Cosmology, Department of Astronomy, University of Science and Technology of China, Hefei 230026, China}
\affil{School of Astronomy and Space Sciences, University of Science and Technology of China, Hefei, 230026, China}

\author{Brian C. Lemaux}
\affil{Department of Physics and Astronomy, University of California, Davis, One Shields Ave., Davis, CA 95616, USA}

\author{Lori M. Lubin} 
\affil{Department of Physics and Astronomy, University of California, Davis, One Shields Ave., Davis, CA 95616, USA}

\author{Olga Cucciati}
\affil{INAF - Osservatorio di Astrofisica e Scienza dello Spazio di Bologna, via Gobetti 93/3 - 40129 Bologna - Italy}

\author{Olivier Le Fèvre}
\affil{Aix-Marseille Univ, CNRS, CNES, Laboratoire d'Astrophysique de Marseille, Marseille, France}

\author{Guilin Liu}
\affil{CAS Key Laboratory for Research in Galaxies and Cosmology, Department of Astronomy, University of Science and Technology of China, Hefei 230026, China}
\affil{School of Astronomy and Space Sciences, University of Science and Technology of China, Hefei, 230026, China}

\author{Wenjuan Fang}
\affil{CAS Key Laboratory for Research in Galaxies and Cosmology, Department of Astronomy, University of Science and Technology of China, Hefei 230026, China}
\affil{School of Astronomy and Space Sciences, University of Science and Technology of China, Hefei, 230026, China}

\author{Debora Pelliccia} 
\affil{UCO/Lick Observatory, Department of Astronomy \& Astrophysics, UCSC, 1156 High Street, Santa Cruz, CA, 95064, USA}

\author{Adam Tomczak} 
\affil{Department of Physics and Astronomy, University of California, Davis, One Shields Ave., Davis, CA 95616, USA}

\author{John McKean}
\affil{Kapteyn Astronomical Institute, University of Groningen, Groningen, the Netherlands}

\author{Neal A. Miller}
\affil{Stevenson University, Department of Mathematics and Physics, 1525 Greenspring Valley Road, Stevenson, MD, 21153, USA}

\author{Christopher D. Fassnacht}
\affil{Department of Physics and Astronomy, University of California, Davis, One Shields Ave., Davis, CA 95616, USA}

\author{Roy Gal}
\affil{University of Hawai'i, Institute for Astronomy, 2680 Woodlawn Drive, Honolulu, HI 96822, USA}

\author{Denise Hung} 
\affil{University of Hawai'i, Institute for Astronomy, 2680 Woodlawn Drive, Honolulu, HI 96822, USA}

\author{Nimish Hathi}
\affil{Space Telescope Science Institute, 3700 San Martin Drive, Baltimore, MD 21218, USA}

\author{Sandro Bardelli}
\affil{INAF - Osservatorio di Astrofisica e Scienza dello Spazio di Bologna, via Gobetti 93/3 - 40129 Bologna - Italy}

\author{Daniela Vergani}
\affil{INAF - Osservatorio di Astrofisica e Scienza dello Spazio di Bologna, via Gobetti 93/3 - 40129 Bologna - Italy}

\author{Elena Zucca}
\affil{INAF - Osservatorio di Astrofisica e Scienza dello Spazio di Bologna, via Gobetti 93/3 - 40129 Bologna - Italy\\Received 2021 January 7; revised 2021 March 4; accepted 2021 March 11; published 2021 May 5}

\begin{abstract}

Radio Active Galactic Nuclei (RAGNs) are mainly found in dense structures (i.e., clusters/groups) at redshifts of z$<$2 and are commonly used to detect protoclusters at higher redshift. Here, we attempt to study the host and environmental properties of two relatively faint ($\mathrm L_\mathrm{1.4GHz} \sim10^{25}$ W Hz$^{-1}$) RAGNs in a known protocluster at z=3.3 in the PCl J0227-0421 field, detected using the latest radio observation obtained as part of the Observations of Redshift Evolution in Large-Scale Environments (ORELSE) Survey.  Using new spectroscopic observations obtained from Keck/MOSFIRE as part of the Charting Cluster Construction with the VIMOS Ultra-Deep Survey (VUDS) and ORELSE (C3VO) survey and previous spectroscopic data obtained as part of the VIMOS-VLT Deep Survey (VVDS) and VUDS, we revise the three-dimensional overdensity field around this protocluster. The protocluster is embedded in a large scale overdensity protostructure. This protostructure has an estimated total mass of $\sim$2.6$\times10^{15} M_\odot$ and contains several overdensity peaks. Both RAGNs are hosted by very bright and massive galaxies, while their hosts show extreme differences color, indicating that they have different ages and are in different evolutionary stages.  Furthermore, we find that they are not in the most locally dense parts of the protostructure, but are fairly close to the centers of their parent overdensity peaks. We propose a scenario where merging might already have happened in both cases, which lowered the local density of their surrounding area and boosted their stellar mass. This work is the first time that two RAGNs at low luminosity have been found and studied within a high redshift protostructure.

\end{abstract}


\section{Introduction} \label{sec:intro}

Galaxy clusters and superclusters provide an excellent laboratory for investigating astrophysical phenomena, such as the evolution of galaxies and the dynamics and content of the universe by constraining on cosmological parameters. 
In the local and intermediate ($z \le 1.5$) universe, the relationship between environment and galaxy evolution has been well studied and has converged to a somewhat coherent picture at least on some aspects of this relationship. In general, the galaxies populating structures in the low-redshift universe have come to the end of their evolution, and those populating structures in the intermediate redshift have a higher fraction of star-forming, bluer, and late-type galaxies \citep{Cooper2007, Peng2010, Lemaux2012, Muzzin2013, Cucciati2017, Tomczak2019, Lemaux2019, Old2020, VanderBurg2020}. However, this relationship remains an open question in the high redshift universe. 
It is challenging not only to homogeneously measure environments and systemically search for protoclusters, due to the small number of galaxies, but also to confirm galaxies in those structures without biases on the galaxy population (see review paper \citealp{Overzier2016}).
There is an increasing number of surveys using various methods to search for protoclusters or high-redshift clusters, such as surveying galaxies around quasars or radio galaxies (e.g., \citealp{Venemans2007, Overzier2008, Wylezalek2013, Wylezalek2014, Cooke2015, Cooke2016}), using Ly$\alpha$ emitters or dropout galaxies to trace large-scale structures (e.g., \citealp{Steidel1998, Toshikawa2016, Higuchi2019, Toshikawa2018, Hu2021}), searching for high-redshift overdensities using large deep photometric and spectroscopic surveys (e.g., \citealp{Lemaux2014a, Cucciati2018, Lemaux2018, Toshikawa2020, Darvish2020}), and selecting via the Sunyaev-Zel'dovich (SZ) effect signature (e.g., \citealp{Bleem2015, Everett2020}).

Among these methods, Radio Active Galactic Nuclei (RAGNs) are commonly used as beacons associated with massive galaxies \citep{Overzier2009} and protoclusters in the early universe \citep{Pentericci2000, Venemans2002, Rottgering2003, Miley2004, Hatch2011a}, as observational studies show that RAGNs statistically lie in denser environments compared to radio-quiet galaxies of the same stellar mass (e.g., \citealp{Wylezalek2013, Wylezalek2014, Hatch2014, Malavasi2015}). Simulations also predict that RAGNs are associated with progenitors of massive clusters (e.g., \citealp{Orsi2016}). 
These RAGNs are mostly selected from large radio surveys which have a limitation on the detection of fainter radio sources. Thus, these studies might only reveal the most powerful radio sources ($\mathrm{L_{1.4GHz} \ge 10^{26} W\ Hz^{-1}}$) that are associated with overdense environments. However, less powerful high-z radio galaxies that are likely be missed in such survey, may show differences in their host and environmental properties from their more powerful counterparts. 
This has been found in the low- to intermediate-redshift studies (z $\leq$ 1.5) that are able to detect RAGNs at a lower radio power threshold. Radio galaxies that are moderately powerful in terms of radio luminosity (L$_\mathrm{1.4GHz} \geq 10^{24}$ W Hz$^{-1}$), though generally less powerful than the high-z RAGNs discussed above, are found to be hosted by red and quiescent galaxies and preferentially reside in the cores of clusters and dense regions (e.g., \citealp{Best2005, Shen2017, Magliocchetti2018b}). Conversely, fainter radio galaxies are mostly galaxies undergoing episodes of star formation (e.g., \citealp{Bonzini2015, Smolcic2009a, Padovani2016}). They do not show a clear environmental preference to be in dense environments (e.g., \citealp{Miller2002, Best2004, Shen2017}). 
In fact, even for high radio power high-redshift RAGNs, only $\sim$55\% of RAGNs with $L_\mathrm{500 MHz} \geq 10^{27.5}$ W Hz$^{-1}$ are found to reside in overdense regions at the 2$\sigma$ level \citep{Wylezalek2013}. 

In this study, we investigate the host and environmental properties of two moderately powerful ($\mathrm{L_{1.4GHz} \sim 10^{25} W\ Hz^{-1}}$) RAGNs in the known protocluster PCl J0227-0421 at z=3.29 in the CFHTLS-D1 field \citep{Lemaux2014a}. This protocluster was originally confirmed by both a significant overdensity in photometric redshift members and spectroscopically confirmed members. Combining new spectral data, we revise the three-dimensional (3D) density map of the protocluster. The protocluster is embedded in a large scale overdensity protoprotostructurestructure around the original detection with several clear overdensity peaks connected by filaments. We make use of the Jansky Very Large Array (JVLA) 1.4-GHz imaging obtained as part of the Observations of Redshift Evolution in Large Scale Environments Survey (ORELSE, \citealp{Lubin2009}). Given the deep radio imaging, we search for radio sources in this protostructure, complete down to $\sim$50 $\mu$Jy, which corresponds to L$_\mathrm{1.4GHz} \sim 10^{25}$ W Hz$^{-1}$ at z = 3.3.  

This paper is laid out as follows. Section \ref{sec:obs} provides an overview of the spectroscopy, optical imaging, and radio data available in the CFHTLS-D1 field, as well as the method used to derive the physical parameters of galaxies and their environment measurements. Section \ref{sec:hosts} and \ref{sec:env} describe the host and environmental properties of the two RAGNs.
In section \ref{sec:discussion} we discuss the evolution of protostructure and the possible role of the environment in the formation of RAGNs. 
Finally, section \ref{sec:sum} presents a summary of our results. Throughout this paper, all magnitudes, including those in the IR, are presented in the AB system \citep{Oke1983, Fukugita1996}. We adopt a standard concordance Lambda cold dark matter ($\Lambda$CDM) cosmology with $H_0$ = 70 km s$^{-1}$, $\Omega_\Lambda$ = 0.73, and $\Omega_\mathrm{M}$ = 0.27.

\section{DATA AND METHODS} \label{sec:obs}

\subsection{The Protocluster Cl J0227-0421} \label{sec:proto}

Over the past two decades, the CFHTLS-D1 field has been the subject of exhaustive photometric and spectroscopic campaigns. First observed in broadband imaging as one of the fields of the VIMOS VLT Deep Survey \citep{LeFevre2004}, this field was subsequently adopted as the first of the ``Deep'' fields (i.e., D1) of the Canada-France-Hawaii Telescope Legacy Survey (CFHTLS)\footnote{\url{http://www.cfht.hawaii.edu/Science/CFHTLS/}}. Intensive spectroscopic observations were then taken as part of the VIMOS Ultra-Deep Survey (VUDS; \citealp{LeFevre2015}). 
The primary goal for the survey is to measure the spectroscopic redshifts of a large sample of galaxies at redshifts 2 $\lesssim$ z $\lesssim$ 6. 
Using these spectroscopic data, \citet{Lemaux2014a} performed a systematic search for overdensity environments in the early universe (z $>$ 2) and discovered a massive protocluster PCl J0227-0421 at z = 3.29. By analyzing the spectra, three broadline Type-1 AGN are identified, with one of them being the Brightest Cluster Galaxy (BCG)  in the protocluster and detected as an X-ray AGN. No radio member galaxies were reported, due to the shallow depth of radio observations \citep{Lemaux2014a}.  

\subsection{Spectroscopic data} \label{sec:spec}

\subsubsection{Surveys with VLT/Visible Multi-object Spectrograph (VIMOS)} \label{sec:spec_vuds}

The primary spectroscopic data available in the CFHTLS-D1 field come from VIsible MultiObject Spectrograph (VIMOS, \citealp{LeFevre2003}) taken as part of the VIMOS Ultra-Deep Survey (VUDS; \citealp{LeFevre2015}) and the VIMOS-VLT Deep Survey (VVDS, \citealp{LeFevre2005, LeFevre2013}). These observations have been fully described in \citet{Lemaux2014a}. 

Spectroscopic redshifts of these targets were assessed and fully discussed in \citet{LeFevre2015}. We adopt the reliability thresholds for secure spectroscopic redshifts, following \citet{LeFevre2013}, that the probability of the redshift being correct in excess of 75\% are considered reliable (hereafter ``secure spectroscopic redshifts''). Those objects having secure spectroscopic redshifts are flagged to X2, X3, X4 \& X9, where X=0-3\footnote{X=0 is reserved for target galaxies, X=1 for broadline AGN, X=2 for non-targeted objects that fell serendipitously on a slit at a spatial location separable from the target, and X=3 for those non-targeted objects that fell serendipitously on a slit at a spatial location coincident with the target. For more details on the probability of a correct redshift for a given flag, see \citet{LeFevre2015} and Appendix B of \citet{Lemaux2020}.}. 
The spectroscopic data of one of the newly-confirmed RAGN hosts is taken from the VVDS survey. Its spectrum is shown in Appendix \ref{app:sed} Figure \ref{fig:spec_protoBCG} (also see Figure 11 of \citealt{Lemaux2014a}) .

\begin{figure}
	\centering
	\includegraphics[width=\columnwidth]{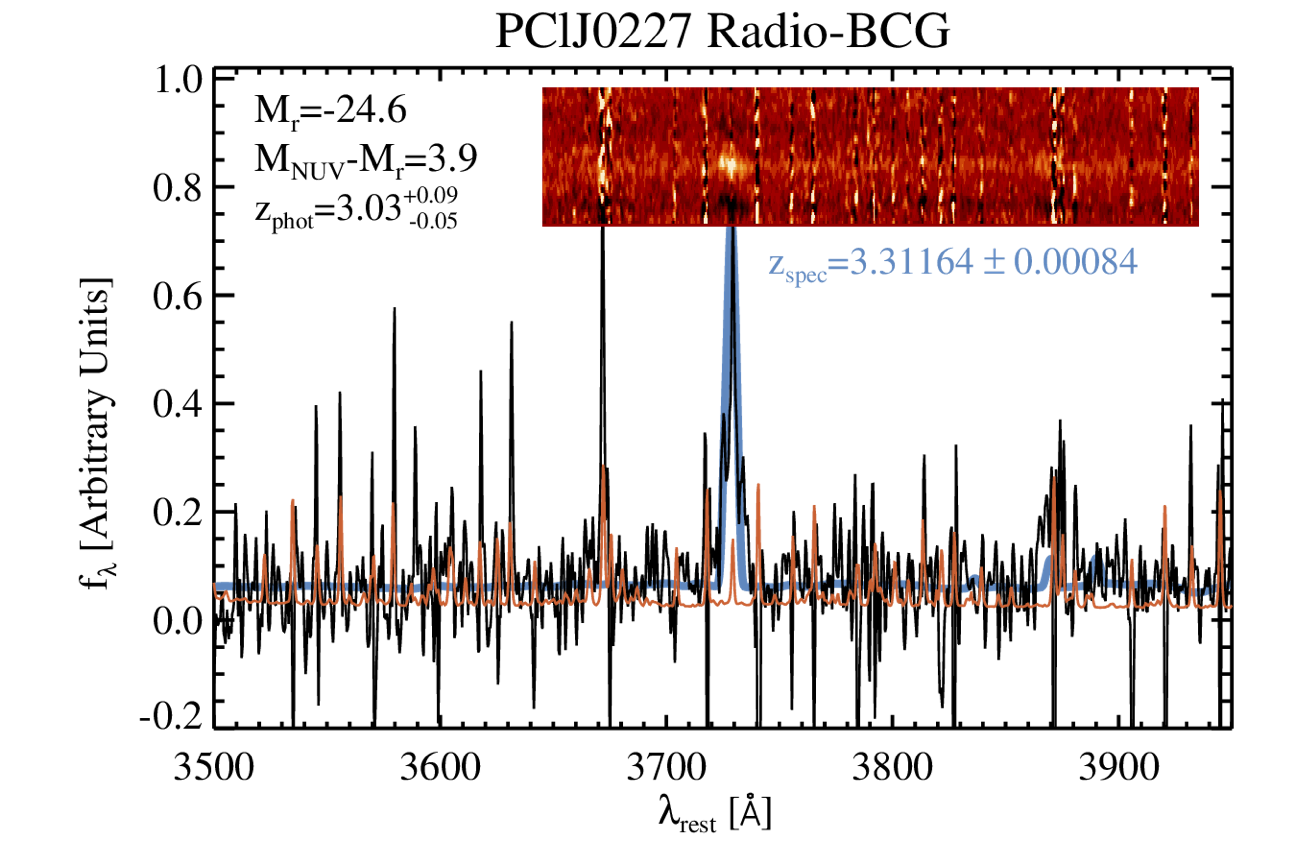}
	\caption{An example of one of the galaxies in the MOSFIRE observation: the 2d/1d spectrum of the radio-BCG. The black line is the one-dimensional flux density spectrum, the orange line is the formal uncertainty spectrum, the blue line is the best-fitted model (see \citealp{LeFevre2015} and references therein for details on the generation of the uncertainty spectrum). The SED best-fitted results for this galaxy are shown on the top left corner. The top inset panel shows the two-dimensional MOSFIRE spectrum. The best-fitted spectroscopic redshift and associated error is given below the inset panel. The spectrum shows a strong [OII] emission (3727\AA) along with a hint of a strong [NeIII] emission (3869\AA) overlapped by a skyline. \label{fig:spec}}
\end{figure}

\subsubsection{The C3VO Survey} \label{sec:spec_c3vo}

Here we describe new spectroscopic data that were taken with the Multi-Object Spectrometer For Infra-Red Exploration (MOSFIRE; \citealt{McLean2012}) on the Keck I 10-m telescope as part of the Charting Cluster Construction with VUDS and ORELSE (C3VO) survey. The Keck component of the C3VO survey is designed to use MOSFIRE and the DEep Imaging Multi-Object Spectrograph (DEIMOS; \citealt{Faber2003}) on Keck I/II, respectively, in order to provide a nearly complete mapping of the five most significant overdensities detected in the VUDS fields. These include overdensities reported in \cite{Lemaux2014a}, \cite{Cucciati2014}, \cite{Lemaux2018}, and \cite{Cucciati2018}. 
This mapping is performed by targeting star-forming galaxies to $i_{AB}<25.3$ or $\sim L^{\ast}_{FUV}$ at $z\sim4.5$ and $<L^{\ast}_{FUV}$ at $z\sim2.5$, where $L^{\ast}_{FUV}$ is the characteristic luminosity measured at $\sim$1600\AA~ rest-frame (see, e.g., \citealp{Bouwens2015, Fuller2020}), and Lyman-$\alpha$ (hereafter Ly$\alpha$) emitting galaxies to fainter magnitudes. Additionally, MOSFIRE allows the targeting of quiescent and dusty star-forming candidates, something that is not possible with DEIMOS, as MOSFIRE observations probe the rest-frame near-ultraviolet (NUV) to optical where such galaxies can feasibly be probed with spectroscopy. 

In total, three MOSFIRE masks were targeted in and around PCl J0227-0421. These three masks were centered at [$\alpha_{J2000}$, $\delta_{J2000}$] = [\mbox{02:27:05.21}, \mbox{-04:19:26.1}] (mask1), [\mbox{02:27:03.5}, \mbox{-04:24:20.0}] (mask2), and [\mbox{02:26:43.5}, \mbox{-04:21:11}] (mask3), with position angles of -21$^{\circ}$,110$^{\circ}$, and 60$^{\circ}$, respectively. Targets for each mask were selected through a series of photometric-redshift and magnitude criteria that largely followed those described in \cite{Lemaux2020}. Briefly, multiwavlength (i.e., X-ray, radio, or mid-infrared detected) objects that lacked a secure redshift were given the highest priority, followed by optical/NIR targets that lacked a secure spectroscopic redshifts but had a photometric redshifts which was in proximity to the adopted redshift range of the protocluster or the surrounding structure. These optical/NIR targets had their priority weighted by the percentage of their reconstructed probability density function (PDF) that fell within the adopted redshift bounds of the protocluster ($3.26 \le z \le 3.35$). At every priority level, we broke the potential target pool into a brighter ($H_{AB}<23.5$) and fainter ($23.5\la H_{AB}<24.5$) sample and up-weighted the priority of brighter potential targets. We also included known structure members with secure spectral redshifts as potential targets as the rest-frame NUV/optical MOSFIRE spectra are complementary to the rest-frame far-UV from VVDS and VUDS. Additionally, fainter targets were allowed to be placed on multiple masks. The MAGMA\footnote{\url{https://www2.keck.hawaii.edu/inst/mosfire/magma.html}} software was used to design all three masks. 

These masks were observed December 17$^{th}$ and 19$^{th}$, 2018 (UTC) under clear skies and seeing that ranged from 0.6$\arcsec$-1.06$\arcsec$, with an average seeing of $\sim$0.75$\arcsec$. Observations were taken in the $H$ band and configurable slit units (CSUs) were configured to employ 0.7$\arcsec$-wide slits. This configuration resulted in an $R\sim3500$. Observations were taken in ``mask nod'' mode, with a nod length of 1.25$\arcsec$. Total integration times were 2.5, 2.4, and 1.9 hours for mask1, mask2, and mask3, respectively. Standard flat-field frames were used for calibration and flux calibration was performed using observations of the standard star HIP13917. 

All raw data were reduced with the \emph{python}-based \texttt{MOSFIRE Data Reduction Pipeline (DRP)}\footnote{\url{http://www2.keck.hawaii.edu/inst/mosfire/drp.html}}, which provided dark-subtracted, flat-fielded, rectified, wavelength-calibrated, background-subtracted two-dimensional flux density and variance arrays for every slit. A custom set of \textsc{IDL} packages was used to collapse all two-dimensional flux density spectra output by the pipeline along their dispersion axis at or near the predicted MAGMA location. A Gaussian was iteratively fit to the resulting collapsed profile using the MAGMA location as the initial guess for the mean spatial location of the targeted galaxy. The final parameters of the Gaussian fit, mean and $\pm$1.5$\sigma$, set the limits on the boxcar extraction used to generate the one-dimensional flux density and noise spectrum. In cases where the continuum was marginally detected in MOSFIRE or only emission lines were present, the dispersion axis would be collapsed over a limited wavelength range and the fit was done by hand. Additionally, all two-dimensional spectra were visually inspected by eye to identify serendipitous detections and all such detections were extracted in a manner identical to target galaxies. Redshifts were determined by using a custom \textsc{IDL} redshift fitting routine based on the Deep Evolutionary Extragalactic Probe 2 (DEEP2; \citealt{Davis2003, Newman2013}) \emph{spec1d} software that relies on $\chi^2$ minimization to a set of empirical templates. In total, 27 unique objects from the three masks yielded secure spectroscopic redshifts including 11 that had redshifts within the adopted redshift range of the protostructure ($3.26 \le z \le 3.35$). 
An example of one of the galaxies in the MOSFIRE observations, which is a newly-confirmed RAGN in the protostructure, is shown in figure \ref{fig:spec}. 
The spectrum shows a strong [OII] emission (3727\AA). Since this RAGN appears to be a quiescent galaxy (see Section \ref{sec:hosts}), the [OII] emission is most likely originate from LINER activity that has been found in quiescent galaxies at low to intermediate redshift \citep{Yan2006, Lemaux2010, Lemaux2017}.

\subsection{Imaging and photometry} \label{sec:photo}

The construction of the existing imaging data and photometric parameters adopted in this paper are described in \citet{Lemaux2014a}. Briefly, deep five-band ($u^{\ast}$ g$^{\prime}$ r$^{\prime}$ i$^{\prime}$ z$^{\prime}$) optical imaging of the entire CFHTLS-D1 field was taken with Megacam \citep{Boulade2003} as part of the ``Deep'' portion of the CFHTLS survey. Approximately 75\% of CFHTLS-D1 field, including the entire area of interest for the present study, was imaged with WIRCam \citep{Puget2004} in the near infrared (NIR) J, H, and Ks bands as part of the WIRCam Deep Survey (WIRDS; \citealp{Bielby2012}). In addition, a large portion of the CFHTLS-D1 field was imaged at 3.6/4.5/5.8/8.0 $\mu$m from the Spitzer InfraRed Array Camera (IRAC; \citealp{Fazio2004}) and at 24$\mu$m from the Multiband Imaging Photometer for Spitzer (MIPS; \citealp{Rieke2004}) as part of the Spitzer Wide-Area InfraRed Extragalactic survey (SWIRE; \citealp{Lonsdale2003}). For further discussion of the available imaging and its depth see \citet{Lemaux2014a, Lemaux2014b}. 

Spectral Energy Distribution (SED) fitting was performed on the observed-frame optical/NIR broadband photometry to estimate photometric redshifts ($z_\mathrm{phot}$), restframe color, stellar masses as well as other properties of the stellar populations of galaxies, utilizing the Le Phare package\footnote{\url{https://www.cfht.hawaii.edu/~arnouts/LEPHARE/lephare.html}} \citep{Arnouts1999, Ilbert2006, Ilbert2009}. The construction of the existing data and photometric parameters adopted in this paper are fully described in \citet{Lemaux2014a}. 
For galaxies with secure spectroscopic redshifts, including those newly observed, the SED fitting process is performed using an identical methodology to other objects with secure spectroscopic redshifts. 
In addition, we visual examine the best-fitting SEDs of RAGN hosts. We find a good agreement between the models and the observed photometry with no strong rise in the IR bands. As a check, we recover statistically identical stellar mass if bands red-ward of Ks are excluded in the fitting. Thus, we confirm that the reported stellar mass are representative of these host galaxies despite the presence of non-stellar emission. 
The observed photometry along with the SED fits of the two RAGN hosts are shown in the Appendix \ref{app:sed} Figure \ref{fig:sed}, and clearly demonstrate the extremely different properties of the host galaxies of the two RAGNs. 

\subsection{Sample Selection} \label{sec:sample} 

For our sample selection, we want to fully investigate the properties of RAGNs as compared to galaxies over a wide range of local and global environments in the same redshift range and the role of environment on the formation of radio emission. We, therefore, select a coeval galaxy sample that includes galaxies that are in and near the protostructure, while minimizing the number of field galaxies near the protostructure that might dilute any signal of the protocluster. 
Therefore, we carefully select a three-dimensional (3D) box with 36.50 $\leq$RA $\leq$ 36.85 and -4.50 $\leq$ Decl. $\leq$ -4.20, corresponding to a proper distance of 10 $\times$ 8 Mpc, and $3.26 \le \mathrm{z}_\mathrm{spec} \le 3.35$ ($\sim$19 pMpc or $\sim$6000 km s$^{-1}$) along the line of sight (l.o.s.). 
For galaxies without a secure spectral redshift, we extend the redshift range to 2.77 $\leq$ z$_\mathrm{phot}$ $\leq$ 3.86 given by the 1.5$\sigma_{z/(1+z)}$ of z$_\mathrm{spec}$ range with $\sigma_{z/(1+z)} = 0.08$, to account for the uncertainty of z$_\mathrm{phot}$\footnote{$\mathrm{z_{phot,min} = z_{spec,min} - 1.5\sigma \times (1 + z_{spec,min})}$ \\ $\mathrm{z_{phot,max} = z_{spec,max} + 1.5\sigma \times (1 + z_{spec,max})}$}. 
The $\sigma_{z/(1+z)}$ is estimated as the $\sigma_\mathrm{NMAD}$ from a comparison between the spec-$z$ and photo-$z$ for objects with flags = X2, X3, X4, \& X9 (see the flag definition in Section \ref{sec:spec_vuds}), and in the redshift range 3 $<\mathrm{z}_\mathrm{spec}<$ 4. Note that we exclude galaxies/stars that have a secure spectral redshift outside this range in the photo-$z$ sample. 

Photometric objects and spectroscopically-confirmed galaxies that reside in these three-dimensional boxes are selected as ``coeval photo-$z$ objects'' and ``coeval spec-$z$ galaxies''. 
In addition, a stellar mass cut $M_* \ge10^9$ $M_\odot$ is imposed for both samples due to severe incompleteness at these masses from both an imaging and spectroscopic perspective (see more discussion in \citealp{Lemaux2020}).  
We obtain a total sample of 48 coeval \mbox{spec-$z$} galaxies, including 19 new spectroscopically-confirmed galaxies obtained from Keck/MOSFIRE. We note that additional five spec-$z$ galaxies in the selected 3D box were excluded with their stellar mass $M_* < 10^9$ $M_\odot$. 
Our photo$-z$ and stellar mass cuts result in a total sample of 4215 coeval \mbox{photo-$z$} objects. 
In this work, we primarily focus on spectroscopically-confirmed galaxies. The coeval photo-$z$ objects are used only to contextualize the spec-$z$ galaxies.
We note that both coeval spec-$z$ galaxies and photo-$z$ objects are some combination of true protostructure members and coeval field galaxies. However, we do not define precise criteria for ``membership'' in this paper and rather choose the non-binary estimates of global and local environment (see section \ref{sec:env_method}) to assess possible scenarios of environmentally-driven evolution. 

\subsection{Radio observation and the detection of RAGNs} \label{sec:radio}

The radio images were mapped using the Karl G. Jansky Very Large Array (VLA) at 1.4GHz in its B configuration, where the synthesized beam is about 5'' (FWHM) and the field of view (i.e., the FWHM width of the primary beam) is approximate 31' in diameter. The radio final image has a sensitivity of 13.6 $\mu$Jy. The final radio catalogs contain sources above 4$\sigma$ and down to $\sim$50$\mu$Jy. Data reduction and source catalogs for this field are obtained following the same methodology as in \citet{Shen2020a}. 
The original observation was taken as part of the ORELSE survey, targeting the foreground cluster XLSS005 at z $\sim$ 1.05 \citep{Valtchanov2004, Lemaux2019, Tomczak2019, Hung2020}, but the $z\sim3.3$ protostructure that is the subject of this paper is contained in the same field of view. 
We note that we did not combine other, shallower VLA observations of the CFHTLS-D1 field because the extra depth was not necessary to characterize the intermediate-to-high power density RAGNs, which was the goal of this paper. 
The CFHTLS-D1 field has also been imaged at the Giant Millimetre Radio Telescope (GMRT) at 610 MHz, which is discussed in the next section. 

To identify the optical counterparts to these radio sources, we perform a maximum likelihood ratio (LR) technique following section 3.4 in \citet{Rumbaugh2012}. We adopt a primary search radius of $2\arcsec$ for the overall photometric catalogs that have observed H or K band magnitude. These bands are in similar restframe wavelength as observed i'-band at z $\sim$ 1 that was used in previous optical matching in the ORELSE survey (e.g., \citealp{Rumbaugh2012, Shen2020a}).  We run parallel optical matchings using H/K band magnitudes and obtain the same matches. In addition, we run a search with a $5\arcsec$ radius to spot additional matches to sources to account for astrophysical and astrometric offsets for multiple components systems. No additional matches are added in this step. 
We cross match radio matches to coeval photo-$z$ objects/spec-$z$ galaxies, and obtain a sample of four matches to coeval photo-$z$ objects, including two matches to coeval spec-$z$ galaxies. 
The latter two spectroscopically-confirmed matches are also detected in X-ray (see section \ref{sec:Xray}). Thus, they were identified as AGN and we attributed at least part of their radio emission as originating from AGN activity. 

\begin{figure}
	\centering
	\includegraphics[width=\columnwidth]{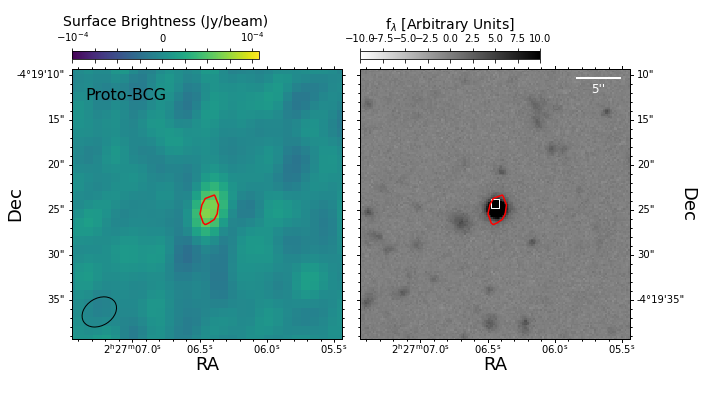}\vspace{-3mm} 
	\vspace{-3mm}
	\includegraphics[width=\columnwidth]{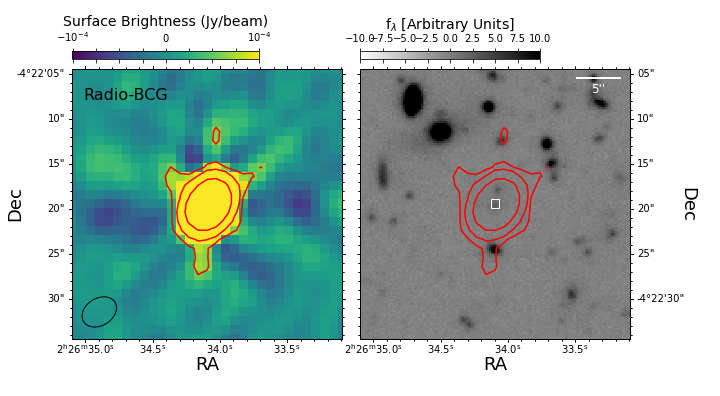}
	\caption{The 1.4 GHz (left) and K-band (right) cutout images of proto-BCG (top) and radio-BCG (bottom), with 4, 16, 64$\sigma$ radio contours overlaid. Their optical counterparts are marked by white boxes. Images are 30'' $\times$ 30'' indicated by the scale bar in the upper right corner. The synthesized beam size is shown in the lower left hand corner.  North is up and east is to the left.  \label{fig:cutouts}}
\end{figure}

One of the RAGNs is hosted by the protocluster galaxy that is the brightest in the optical/NIR, named as ``proto-BCG'' in \citet{Lemaux2014a}. We continue to use this name since, as we will show later, it is still the brightest optical/NIR galaxy in the protostructure. In addition, the proto-BCG is detected in the Spectral and Photometric Imaging REceiver (SPIRE; \citealp{Griffin2010}) aboard the Herschel Space Observatory \citep{Pilbratt2010}.  The SFR of proto-BCG, estimated from the total infrared luminosity, is $750 \pm 70 M_\odot / yr $. 
The second radio galaxy is found to be more radio luminous. Thus, we named this the brightest radio protostructure galaxy ``radio-BCG''. We note that these names are mostly for simplification in the text. The 1.4 GHz and K-band cutout images are shown in Figure \ref{fig:cutouts}. The 1d/2d spectrum of radio-BCG is shown in Figure \ref{fig:spec}. 
The flux densities and radio powers are listed in table \ref{tab:prop}. We present radio powers using three slope indexes ($\alpha$ = -0.77, 0.2 and 0.7), which are further explained in Section \ref{sec:610MHz}. 
     
\begin{deluxetable*}{lcccccccccc}
\tablecolumns{9}
\tablewidth{0pt}
\tablecaption{Radio and X-ray Properties of RAGNs \label{tab:prop}}
\tablehead{
\colhead{Name} & \colhead{RA} & \colhead{Dec} & \colhead{z$_\mathrm{spec}$} & \colhead{$f_\mathrm{1.4GHz}$}& \colhead{log(L$^{\alpha=0.77/0.2/-0.7}_{\mathrm{1.4GHz}}$)}  & \colhead{log(L$^\mathrm{err}_{\mathrm{1.4GHz}}$)}  & \colhead{XMM ID} &\colhead{log(L$_{X}$)} \vspace{-0.2cm} \\
& & & & \colhead{$\mu$Jy} & \colhead{log(W Hz$^{-1}$)} & \colhead{log(W Hz$^{-1}$)}   & &\colhead{log(erg s$^{-1}$)}  \vspace{-0.2cm}\\
\colhead{(1)} & \colhead{(2)} & \colhead{(3)} & \colhead{(4)} & \colhead{(5)} & \colhead{(6)} & \colhead{(7)} & \colhead{(8)} & \colhead{(9)} }
\startdata
proto-BCG & 02:27:06.5 & -04:19:24.3 & 3.2852& 77$\pm$14 & 23.79/24.15/24.72 &  0.15  & XMM04725 & 44.44$\pm$0.09\\ 
radio-BCG & 02:26:34.1 & -04:22:19.4 & 3.3112 & 4304$\pm$16 & 25.53/25.89/26.46 & 0.01 & XMM04484 & 45.18$\pm$0.04 \\
\enddata
\tablecomments{\footnotesize{Column 1-4 are name, coordinates and redshift of RAGNs. Column 5 is the flux density at 1.4GHz. Following \citet{Shen2017}, the peak flux density is used unless the integrated flux is larger by more than 3$\sigma$ than the peak flux for each individual source. The peak flux density is used for proto-BCG and the integrated flux is used for radio-BCG. Column 6 is radio powers calculated by three different slope indexes (see Section \ref{sec:610MHz}). Column 9 is the error on radio power. They are dominated by the error on flux, regardless of the slope index. Column 8 is the ID in XMM-LSS \citep{Chen2018}. Column 9 is the X-ray power calculated using the z$_\mathrm{spec}$. }}
\end{deluxetable*}

\subsubsection{Other radio band cross match} \label{sec:610MHz}

The 610 MHz observation of CFHTLS-D1 field has a depth of $\sim$250 $\mu$Jy (5$\sigma$) and has an angular resolution of $\sim$ 6 arcsec  (see \citealp{Bondi2007} for the more details). 
We crossmatched our RAGNs to the GMRT 610 MHz radio source catalog. Only the radio-BCG is detected with $f_\mathrm{610MHz}$ = 2.26 $\pm$ 0.04 mJy. 
With the two radio bands, we obtain a spectral index $\alpha = 0.77 \pm 0.02$ ($f_\nu \propto \nu^{+\alpha}$). 
Studies have shown that the spectral index for radio sources spans a large range from -2 to +1, which depends on the dominant component of radio emission. A core-dominated source has a flat or inverted spectrum with $\alpha \gtrsim 0.2$, while a jet-dominated source has a steep spectrum with $\alpha \sim -0.7$ (e.g., \citealt{Condon1992, Hovatta2014}). 
However, we are not able to spatially resolve the radio-BCG in either radio observation; thus, it is unknown whether the radio-BCG has small-scale jets. 
In previous studies of radio galaxies in ORELSE, we have assumed a single $\alpha = -0.7$ for all radio galaxies, due to the lack of other radio band observation \citep{Shen2017, Shen2019, Shen2020a, Shen2020b}. As a result, to treat the two RAGNs evenly, we report the radio power at 1.4GHz in Table \ref{tab:prop} assuming three $\alpha$: $\alpha = 0.77$ as we measured for the radio-BCG, a relatively flat spectrum with $\alpha = 0.2$ adopted from the mean value of the core spectral indices \citep{Hovatta2014}, and an $\alpha = -0.7$ as we used in ORELSE. 

 
\subsection{X-ray cross match} \label{sec:Xray}

The XMM-Newton point-source catalog has been constructed for this field as part of the XMM-Large Scale Structure survey region (XMM-LSS) \citep{Chen2018}. 
We run an optical matching to identify X-ray counterparts following the same procedure as described in Section \ref{sec:radio}. 
We found two matches to coeval spec-$z$ galaxies. Interestingly, they are also the RAGNs in the protostructure and the proto-BCG (the latter was already found by \citealp{Lemaux2014a}). This fact potentially indicates that some portion of the radio emission is generated from AGN accretion. We will further discuss the mechanism of radio emission in Section \ref{sec:evolution_RAGN}. 
We calculate their rest-frame 2–10 keV ``apparent'' luminosity using secure z$_\mathrm{spec}$ and assuming a $\Gamma$ = 1.7 power law spectrum corrected for Galactic absorption following the same method as \citet{Chen2018}. These values are listed in Table \ref{tab:prop}. 

\subsection{Environmental Measurements} \label{sec:env_method}

\begin{figure*}
    \centering
    \includegraphics[width=0.8\textwidth]{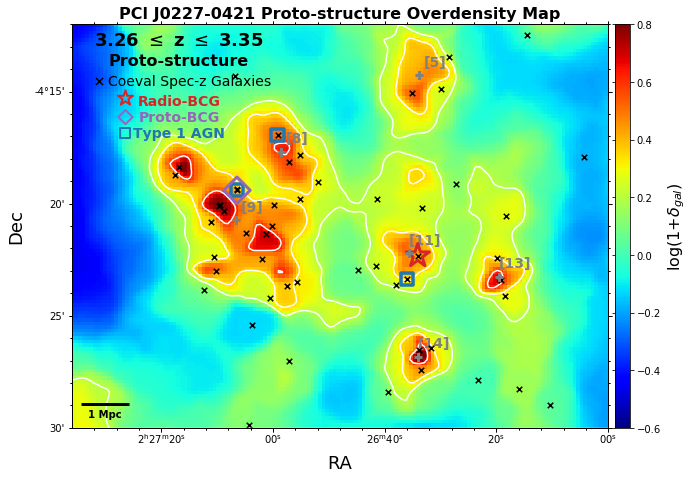}
    \caption{Position of RAGNs (radio-BCG in open red star and proto-BCG in purple diamond), Type-1 AGN (open blue squares) and coeval spec-$z$ galaxies (black crosses) in the 2D overdensity map. The contours represent the 1.5, 2.5, 3.5$\sigma$ overdensity levels with $\sigma$=0.14. The barycenters of the density peaks are marked by grey plus signs with their IDs shown next to it. The peak [9] is the original protocluster. \label{fig:2dmap}}
\end{figure*}

In our z $\sim$ 1 studies of radio galaxies, we have adopted two environmental measurements: a local environment that probes the current density field to which a galaxy is subject and a global environment that probes the time-averaged galaxy density to which a galaxy has been exposed \citep{Shen2017, Shen2020a}. In this paper, we adopt the same environmental measurements but with some modifications. 

The local environment, defined as log(1+$\delta_\mathrm{gal}$), is obtained using a Voronoi tessellation Monte Carlo (VMC) algorithm (see \citealp{Lemaux2018, Cucciati2018, Lemaux2020} for full details). 
In brief, for each redshift slice, a final density map is computed as the median of the density maps among the VMC realisations. $\delta_\mathrm{gal}$ is the local overdensity measured from the final density map with respect to the mean density of that at each redshift. 
This method has been successfully used to search for overdensity candidates at z$\sim$1 in the ORELSE survey \citep{Hung2020} and in the high-redshift ($z>2$) universe \citep{Lemaux2018, Cucciati2018, Lemaux2020}. 
Instead of using a series of multiple redshift slices, we simplify the construction of VMC overdensity map by using a single broad redshift range covering the full protostructure (3.26 $\leq $z $\leq$ 3.35, $v = \pm$3000 km s$^{-1}$), taken the same statistical approach as \citet{Lemaux2020} for spectral objects of different flags. We name it as the 2D overdensity map, hereafter, to distinguish with the 3D overdensity map in Section \ref{sec:evolution_protocluster}. 
The final map generates a log(1+$\delta_\mathrm{gal}$) map from which the log(1+$\delta_\mathrm{gal}$) of individual RAGN and coeval spec-$z$ galaxies/photo-$z$ objects can be made. For the coeval photo-$z$ objects, we assume that they are in the constructed redshift range. 
The 2D overdensity map is shown in Figure \ref{fig:2dmap} with coeval spec-$z$ galaxies, RAGNs and Type-1 AGNs marked. The contours represent the 1.5, 2.5, 3.5$\sigma$ overdensity levels with $\sigma$=0.14. 

To identify the density peaks, we use standard photometry software package \textsc{SExtractor} \citep{Bertin1996} on the 2D overdensity map using the parameters \textsc{DETECT\_THRESH} of 2.5$\sigma$ and a \textsc{DETECT\_MINAREA} of 5 pixels. The 2.5$\sigma$ is chosen, since regions enclosed by 2.5$\sigma$ contours are dense regions, as shown in the Figure \ref{fig:2dmap}. We identify 6 density peaks in the box defined in Section \ref{sec:sample}. In Figure \ref{fig:2dmap}, the barycenter of density peaks are marked by grey plus signs with their ID given by the side. We keep the ID from SExtractor that is run on the full 2d map, although we are only interested in the surrounding region of the protocluster. The original protocluster is identified as peak 9. Coeval spec-$z$ galaxies within the 2.5$\sigma$ contour of each density peak are identified as spec-$z$ members. The effective radius (R$_\mathrm{eff}$) of each density peak is derived from its 2.5$\sigma$ contour. The barycenter, the number of spec-$z$ members, the median redshift (z$_\mathrm{med}$) of spec-$z$ members and R$_\mathrm{eff}$ of each density peak are listed in Table \ref{tab:peaks}.

In the studies of ORELSE, the global environment is defined as $\eta = R_\mathrm{proj} / R_{200} \times |\Delta v|/\sigma_v$ following the method described in \citet{Shen2019}. However, the density peaks are likely not virialized, and the small number of spec-$z$ galaxies in each density peak prevent a robust estimation of the l.o.s velocity dispersion. Therefore, we only consider the transverse part. As a result, the global environment is simplified as R$_\mathrm{proj, norm}$=R$_\mathrm{proj}$/R$_\mathrm{eff}$, where R$_\mathrm{proj}$ is the project distance of galaxies to the barycenter of their closest peak, and R$_\mathrm{eff}$ is the effective radius defined in the 2D overdensity detection. R$_\mathrm{eff}$ is adopted as a proxy for R$_{200}$ to uniformly compare among these density peaks. We measure the R$_\mathrm{proj, norm}$ for RAGNs, coeval spec-$z$ galaxies and photo-$z$ objects. 

\begin{deluxetable}{ccccccc}
\tablecolumns{7}
\tablewidth{0pt}
\tablecaption{Properties of density peaks from 2D overdensity map \label{tab:peaks}}
\tablehead{
\colhead{ID} & \colhead{RA$_{2D}$} & \colhead{Dec$_{2D}$} & \colhead{n$_{zs}$} &\colhead{z$_\mathrm{med}$} & \colhead{$\langle\delta_{gal,2D}\rangle$} & \colhead{R$_\mathrm{eff}$}  \vspace{-0.2cm}\\
 &  &  & & & & \colhead{(cMpc)}  \vspace{-0.2cm}\\
\colhead{(1)} & \colhead{(2)} & \colhead{(3)} & \colhead{(4)} & \colhead{(5)} & \colhead{(6)} & \colhead{(7)}  }
\startdata
5 & 2:26:33.9 & -4:14:17.2 & 1 & 3.269 & 1.68 & 2.42 \\
8 & 2:26:58.5 & -4:17:41.6 & 3 & 3.325 & 1.88 & 2.95  \\
9 & 2:27:06.6 & -4:20:44.5 & 14 & 3.299 & 2.08 & 5.07 \\
11 & 2:26:35.2 & -4:22:13.1 & 2 & 3.300 & 1.68 & 2.00  \\ 
13 & 2:26:19.6 & -4:23:15.7 & 2 & 3.298 & 2.17 & 1.65 \\ 
14 & 2:26:34.0 & -4:26:49.9 & 2 & 3.302 & 2.17 &1.78 \\
\enddata
\tablecomments{\footnotesize (1) ID of the subcomponent; (2) and (3) are the RA and Dec of the barycenter of the density peaks; (4) is the number of spectroscopic members; (5) is the median redshift of spectroscopic members; (6) and (7) are the average $\delta_{gal, 2D}$ and the circularized transverse radius identified in 2D overdensity map. }
\end{deluxetable}

\section{Host properties of RAGNs} \label{sec:hosts}

\begin{figure*} 
	\centering
	\includegraphics[width=\textwidth]{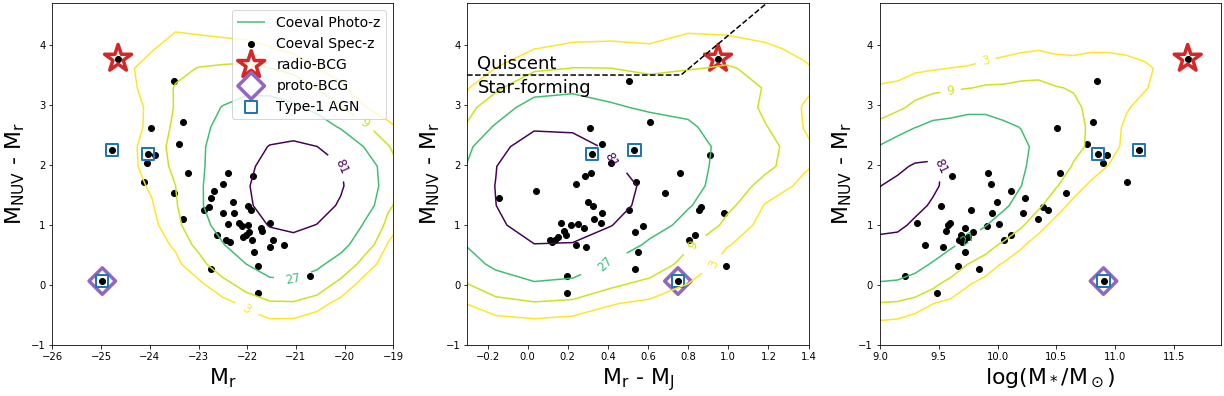}
	\caption{The rest-frame $M_\mathrm{NUV}$ - $M_\mathrm{r}$ versus $M_\mathrm{r}$ (\textit{left}), $M_\mathrm{NUV}$ - $M_\mathrm{r}$ versus $M_\mathrm{r}$ - $M_\mathrm{J}$ (\textit{middle}), and $M_\mathrm{NUV}$ - $M_\mathrm{r}$ versus log($M_*$/$M_\odot$) (\textit{right}) phase space diagrams. The black dots are coeval spec-$z$ galaxies. The colored contours are derived based on the number density distributions of coeval photo-$z$ objects. The numbers of photo-$z$ objects are labeled on the contours. The radio-BCG, proto-BCG and Type-1 AGNs are marked with red open stars, purple open diamonds and blue open squares. The solid black lines in the color-color diagram is the divisions between the SF and quiescent galaxy populations adopted from \citet{Lemaux2014b} at 2 $\le$ z $\le$ 4. All coeval spec-$z$ galaxies are in the SF region. \label{fig:host}}
\end{figure*}

To study the host properties of our RAGNs, we compare the brightness, color, and stellar mass of RAGNs to the coeval photo-$z$ and spec-$z$ galaxies in Figure \ref{fig:host}.  
One of the most striking features of these plots is that the hosts of RAGNs are far different than the majority of galaxies. Specifically, we see a pronounced gap in their $M_\mathrm{NUV}-M_\mathrm{r}$ colors, and are both very bright and massive, as compared to coeval spec-$z$ galaxies. 
The proto-BCG is extremely bright in the rest-frame r band ($M_\mathrm{r}$ = -25.0)\footnote{All absolute magnitudes are uncorrected for extinction.}, but exhibits extremely blue colors ($M_\mathrm{NUV} - M_\mathrm{r}$ = 0.07). It is located in the star-forming region of the color-color diagram. 
It is consistent with being either the first or second most massive galaxy inhabiting its parent density peak (peak 9) and among the top five most massive galaxies in the entire coeval spec-$z$ sample ($M_*$ = 10$^{10.90 \pm 0.04} M_\odot$). 
On the other hand, the radio-BCG is also very bright in the optical, just 0.3 mags fainter in $M_\mathrm{r}$  ($M_\mathrm{r}$ = -24.7) and exhibits the reddest color ($M_\mathrm{NUV}-M_\mathrm{r}$ = 3.77) of all protostructure members. In the middle panel, the radio-BCG is located in the star-forming region but very close to the quiescent/star-forming separation line. It is the most massive galaxy in the protostructure ($M_* = 10^{11.62 \pm 0.04} M_\odot$). 
The other two Type-1 AGN hosts have similar $M_\mathrm{NUV}-M_\mathrm{r}$ colors, both of which are in between that of RAGN hosts. They are also hosted by very bright and massive galaxies. 

To compensate for the incompleteness of our coeval spec-$z$ galaxy sample, we overlap the 2-dimensional number density contours of coeval photo-$z$ objects in Figure \ref{fig:host}. The color of the contours are lighter for lower number densities. We find that RAGNs also reside away from the majority of coeval photo-$z$ objects.

The position of a galaxy in the color-stellar mass diagram, especially its color, modulo dust, metallicity, and unobscured AGN effects, is a good indicator of the age of a galaxy (i.e., the time since the inception of the star formation event; see trends in Figure 7 in \citealp{Lemaux2018}). 
Regarding AGN activity, for the proto-BCG, being a type-1 AGN host, it is almost certain that the AGN is contributing appreciably to the blueness of its $M_\mathrm{NUV} - M_\mathrm{r}$ color. Conversely, the radio-BCG appears to be a type-2 AGN and the AGN is unlikely to contribute considerably to its UV/optical broadband emission. In the following paragraphs, we combine color with other properties to further discuss the age and possible evolutionary scenario of the two RAGNs. 

The proto-BCG is located in the star-forming region, indicating less time since the inception of the star formation event. In fact, in the previous study, the total infrared luminosity of the proto-BCG implies that it is forming stars at a rate of SFR$_\mathrm{TIR} = 750 \pm 70 \mathrm{M}_\odot \mathrm{yr}^{-1}$ \citep{Lemaux2014a}. 
If we assume that a portion of the radio emission comes from star formation (SF), the expected radio power for a galaxy with that level of star formation is log(L$_{1.4 GHz}$) $\sim$ 24.35 W Hz$^{-1}$, derived using the SFR formula from 1.4 GHz from \citet{Bell2003} and multiplying the derived SFR by a factor of 0.6 to convert from a Salpeter to a Chabrier IMF. 
Here, we adopt the radio power calculated with the typical $\alpha$ = -0.7 (see Table \ref{tab:prop}). The rest of radio emission (log(L$_{1.4 GHz}$) $\sim$ 24.4 W Hz$^{-1}$) would originate from the AGN activity. 

This coeval star formation and AGN activities in the proto-BCG is reminiscent of the unique Hybrid population found in radio galaxies in the ORELSE survey at z$\sim$1 \citep{Shen2017}. 
Hybrids, though selected as an intermediate population between AGNs and SFGs in the classification scheme, were found to have coeval star formation and AGN activity with high accretion efficiency. 
In a further study on the hybrid population, a larger sample (179) of radio- and mid-infrared-detected galaxies was selected at 0.55 $\leq$ z $\leq$ 1.30 \citep{Shen2020a}. 
These galaxies were further sorted into four phases according to their Eddington ratio and SFR. Shen et al. proposed an evolutionary scenario where AGN activities ramp up and down, as stellar mass increases constantly, while SFR decreases dramatically during the AGN activity ramp down phase. 
Even though the proto-BCG is at higher redshift, it has comparable stellar mass to the hybrid population at z$\sim$1. It is plausible that the proto-BCG has experienced a similar co-evolution of AGN and SF activities. 

As for the radio-BCG, we do not detect it with Herschel. In addition, the mean luminosity-weighted stellar ages obtained from the SED fitting is $2.0 \times 10^9$ yr for the radio-BCG and $3.2 \times 10^8$ yr for  the proto-BCG, a $\sim$1.7 Gyr difference. Though there are large random and systematic uncertainties on these age estimates (see, e.g., \citealp{Thomas2017}), this difference is considerable. The Big Bang happens only $\sim$2 Gyr prior to z $\sim$ 3.3, which implies that the radio-BCG began forming stars earlier than 300 Myr after the Big Bang. All of the color and age properties of the radio-BCG suggest that the star formation in the radio-BCG has been quenched or is on its way to be quenched. 

Recently, \citet{Forrest2020} studied a sample of 16 spectroscopically-confirmed massive galaxies (M $> 10^{11}$$M_\odot$) at redshifts of z $>$ 3  selected from the COSMOS-UltraVISTA and XMM-VIDEO fields. 
In fact, our radio-BCG is the target ``XMM-VID3-2293'' in their sample (B. Forrest, private communication). Their measured properties of this galaxy, using the combined spectroscopy and photometry, are consistent with our results in terms of stellar mass, age and identification as a narrow-line AGN. Moreover, the radio-BCG host appears to be one of the oldest galaxies in their sample and their results indicate that its star formation is likely to have been quenched over a fairly rapid time scale ($\sim300$ Myr).  
Overall, the radio-BCG host has similar properties to traditional RAGN hosts, which are typically found to be massive, red and quiescent galaxies (e.g., \citealp{Miller2002, Mauch2007, Kauffmann2008, Malavasi2015, Shen2017}), with their AGN generally powered by inefficient accretion (e.g., \citealp{Best2005, Tadhunter2016}). In summary, the two RAGNs appear to have different ages and are in two clearly different evolutionary stages.  The radio-BCG is likely to be a traditional selected RAGN, while the proto-BCG is likely a hybrid galaxy, defined as having vigorous coeval AGN and SF activity. 

\section{The environmental properties of RAGNs} \label{sec:env}

\begin{figure*} 
	\centering
	\includegraphics[width=\textwidth]{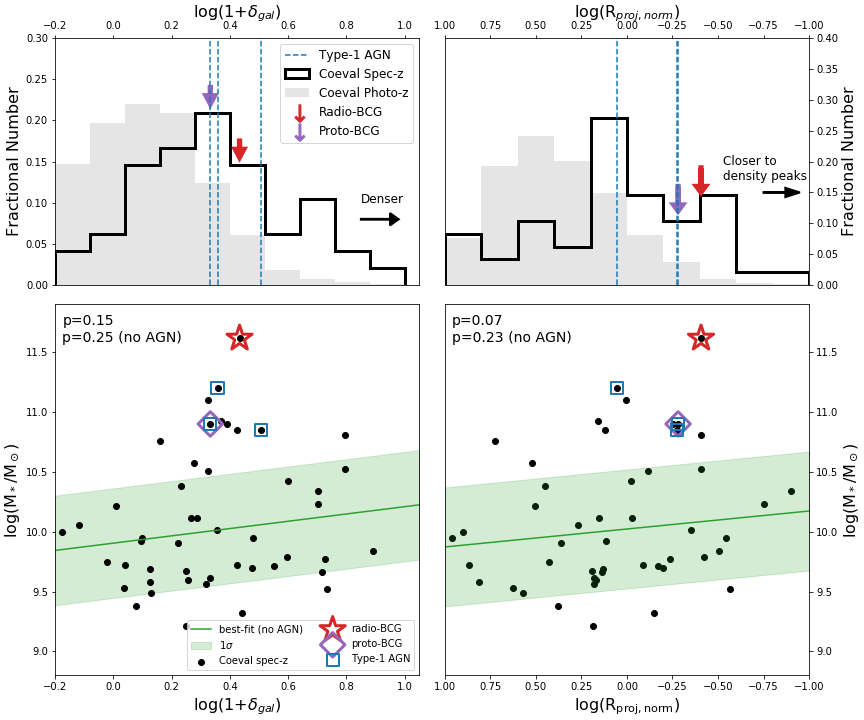}
	\caption{\textit{Top}: Fractional number histograms of the local overdensity log($1+\delta_\mathrm{gal}$) (\textit{left}) and normalized projected distance of galaxies to the center of their parent density peaks R$_\mathrm{proj, norm}$ (\textit{right}) for coeval spec-$z$ galaxies (bold black lines) and photo-$z$ objects (grey shaded). The radio-BCG, proto-BCG and Type-1 AGNs are marked with red arrows, purple arrows and blue vertical lines. \textit{Bottom}: The log($1+\delta_\mathrm{gal}$) (\textit{left}) and R$_\mathrm{proj, norm}$ (\textit{right}) as a function of stellar mass. The black dots are coeval spec-$z$ galaxies. The radio-BCG, proto-BCG and Type-1 AGNs are marked with red open stars, purple open diamonds and blue open squares.  The best-fitted lines of coeval spec-$z$ galaxies, excluding RAGN/Type-1 AGNs, are shown in green solid lines with $1\sigma$ envelopes as shaded regions. The p-value of Spearman tests of coeval spec-$z$ galaxies with and without RAGN/Type-1 AGNs are reported on the upper left corners. \label{fig:env}}
\end{figure*}

The hosts of the two RAGNs in the protostructure are very different in their color properties but are both extremely bright and massive compared to other galaxies at the same redshift.  
In previous studies of RAGNs, RAGNs are preferentially found in the cores of galaxy clusters and locally overdense environments relative to similarly massive radio-undetected galaxies in the local universe (e.g., \citealp{Miller2002, Kauffmann2008}), at intermediate redshift z $\sim$ 1 \citep{Shen2017}, and even at high redshift 1 $\le$ z $\le$ 3 \citep{Hatch2014}. 
On the other hand, the hybrid population, those radio galaxies having coeval AGN and SF activities, does not show clear environmental preferences compared to galaxies of similar color and stellar mass at intermediate redshifts of z$\sim$1 \citep{Shen2017}. 
Here, we see if this trend is borne out for the two RAGN detections in the protostructure surrounding PCl J0227-0421. 

In the top left panel of Figure \ref{fig:env},  the histograms of log($1+\delta_\mathrm{gal}$) for coeval spec-$z$/photo-$z$ populations are shown. The radio-/proto-BCG are marked by purple/red arrows, and Type-1 AGNs are marked by blue dashed lines. 
We find that the RAGNs reside in a moderately dense region with log($1+\delta_\mathrm{gal}$) = 0.33/0.43 for proto-/radio-BCG, respectively. 
The dynamic range of log($1+\delta_\mathrm{gal}$) of coeval spec-$z$ galaxies extents from -0.17 to 0.89. It has a median value of 0.33 with 16\%/84\% values of 0.11/0.60. The log($1+\delta_\mathrm{gal}$) values of RAGNs/Type-1 AGNs are within the 16\%/84\% of that of coeval spec-$z$ galaxies. This naively suggests that there is nothing particularly special about the local environment of the RAGNs relative to that of the average galaxy in the coeval spec-$z$ sample. 

In the top right panel of Figure \ref{fig:env}, the histograms of log(R$_\mathrm{proj, norm}$) of coeval spec-$z$/photo-$z$ populations are shown, and RAGNs and Type-1 AGNs are marked in the same way as the top left panel. 
Both RAGNs are within the effective radius of their parent-peaks and are fairly close to the center of their parent-peaks compared to the coeval spec-$z$ galaxies. 
The log(R$_\mathrm{proj, norm}$) distribution of coeval spec-$z$ galaxies has a median of 0.05 with 16\%/84\% values of -0.39/0.49. The log(R$_\mathrm{proj, norm}$) of radio-/proto-BCG are -0.28/-0.41, at the 20\%/14\% levels of the full distribution, respectively. 
As for the other two Type-1 AGNs, their log(R$_\mathrm{proj, norm}$) are -0.28/0.05, at the 20\%/50\% levels of the full distribution, respectively. Thus, three out of four AGNs reside fairly close to the center of their parent-peaks compared to the coeval spec-$z$ galaxies, i.e., in the lowest quantile of the R$_\mathrm{proj, norm}$ distribution of all coeval spec-$z$ galaxies. 

In the bottom panels of Figure \ref{fig:env}, we show the two environmental measurements of RAGNs, Type-1 AGNs and coeval spec-$z$ galaxies as a function of stellar mass. 
We find that RAGNs and Type-1 AGNs have higher stellar masses than the coeval spec-$z$ galaxies at similar local/global environment. To quantify such an offset, we perform the analysis by fitting a linear relation between the $M_*$ and log($1+\delta_\mathrm{gal}$)/log(R$_\mathrm{proj, norm}$) of the coeval spec-$z$ galaxies excluding RAGNs and Type-1 AGNs, adopting a linear least-squares approach. In the bottom panels of Figure \ref{fig:env}, the best-fitted lines are shown in green with the $1\sigma$ envelops in shaded regions. The $\sigma$ is determined by the standard deviation of the residuals of 100 Monte-Carlo resampling. The proto-/radio-BCG are offset from the best-fitted lines with 2.1$\sigma$/3.7$\sigma$ and 1.8$\sigma$/3.3$\sigma$ significant levels in $\delta_\mathrm{gal}$ and R$_\mathrm{proj, norm}$, respectively. All AGNs are deviate at some level of statistical significance from these relations with higher stellar masses relative to the coeval spec-$z$ galaxies at the same local density and proximity to the density peaks.  

In the local universe, galaxies in higher-density environments have been observed to have higher stellar masses (e.g., \citealp{Kauffmann2004}). At z$\sim$1, \citet{Tomczak2017} using a large sample of eight fields from the ORELSE survey found evidence of environmental effects on the shape of the stellar mass functions. Here, higher local density environments proportionally increases the efficiency of destroying lower mass galaxies and/or growth of higher mass galaxies.  
In a recent study of a large sample of spectroscopically-confirmed galaxies in the early universe (2 $\le$ z $\le$ 5),  \citet{Lemaux2020} found evidence of a weak but highly significant positive correlation between stellar mass and local overdensity. This analysis contained all VIMOS data presented in our study. We adopt the non-parametric Spearman rank correlation test to assess the correlation between stellar mass and local/global density within the much smaller sample presented here. The returned p-value quantifies the significance of the correlation by giving the probability that the data are uncorrelated (i.e., the null hypothesis). We reject the null hypothesis for p-value $\le$ 0.05. 
 The p-values are 0.15/0.07 between the stellar mass and log($1+\delta_\mathrm{gal}$)/log(R$_\mathrm{proj, norm}$) of coeval spec-$z$ galaxies, respectively. The p-values are 0.25/0.23, respectively, when excluding RAGNs and Type-1 AGNs. Given the large p-values, for this protostructure, we do not find any significant correlations between stellar mass and local/global density. We note that, unlike \citet{Lemaux2020}, our result is obtained from a single structure, a much smaller number of galaxies, and a lack of galaxies residing in field-like, low density regions, since most of them reside in the protostructure. However, this lack of significant correlation does not negate the fact that the RAGNs are consistent outliers from these relations. 

We summarize results from the environmental preferences of RAGNs and Type-1 AGNs. All of them reside in moderately dense local environment, and three out of the four reside fairly close to the center of their parent-peaks, compared to coeval spec-$z$ galaxies. They have higher stellar mass than galaxies found at the same local densities and distances from the density peaks. 

It is possible that the observed properties of these host galaxies could be explained by merging activity. Such activity would, first, boost the stellar mass of the host galaxy. In turn, this activity would serve to decrease the number of galaxies in the surrounding area. This decrease in companions would manifests as a drop in the local density, while retaining the proximity to the large-scale density peak. We will further discuss this scenario in Section \ref{sec:evolution_RAGN}.

Furthermore, the radio-BCG resides within the effective radius of the parent density peak and is hosted by a galaxy as massive (10$^{11.62}$ $M_\odot$) as the BCGs found in the low- to intermediate-redshift universe (z $\la$ 1.63) \citep{Lidman2012, Lidman2013, Ascaso2014} that have on average 10$^{11.51}$ $M_\odot$ and 10$^{11.66}$ $M_\odot$ in stellar mass at z$\sim$1 and z$\sim$0, respectively. Its immense mass at z$\sim$3 suggests that the radio-BCG is very likely to be the dominant progenitor of what will be the z$\sim$0 BCG. Such a massive galaxy found in a high-redshift protostructure agrees with recent works showing that the majority of stellar mass buildup happens within BCGs by z = 2 through rapid star formation and early assembly \citep{Ito2019, Rennehan2020, Long2020}. Specifically, \citet{Rennehan2020} used a combination of observationally constrained hydrodynamical and dark-matter-only simulations to show the forward-evolution of a protocluster at z$\sim$4.3. They found the stellar assembly time of 90\% of BCGs is $\sim$370 Myr, corresponding to z $\sim$3.3, similar to the redshift of this work. 

\section{Discussion}
\label{sec:discussion}

\subsection{Three-dimensional structure of the protostructure}
\label{sec:3dmap}

\begin{deluxetable*}{lcccccccccccc}
\tablecolumns{13}
\tablewidth{0pt}
\tablecaption{Properties of density peaks from 3D overdensity map \label{tab:3dpeaks}}
\tablehead{
\colhead{ID} & \colhead{RA$_\mathrm{peak}$} & \colhead{Dec$_\mathrm{peak}$} & \colhead{z$_\mathrm{peak}$} & \colhead{n$_\mathrm{spec}$}& \colhead{$\langle\delta_{gal, 3D}\rangle$} & \colhead{R$_\mathrm{x}$} & \colhead{R$_\mathrm{y}$} & \colhead{R$_\mathrm{z}$}& \colhead{E$_{z/xy}$}& \colhead{Volume} & \colhead{$M_\mathrm{tot}$} & \colhead{$\delta_{corr, gal}$}  \vspace{-0.2cm}\\
 & \colhead{[deg]} & \colhead{[deg]} & & & & \colhead{[cMpc]} & \colhead{[cMpc]} &\colhead{[cMpc]}& &\colhead{[cMpc$^3$]} & \colhead{$10^{14}$$M_\odot$} &  \vspace{-0.2cm}\\
\colhead{(1)} & \colhead{(2)} & \colhead{(3)} & \colhead{(4)} & \colhead{(5)} & \colhead{(6)} & \colhead{(7)} & \colhead{(8)} & \colhead{(9)} & \colhead{(10)} & \colhead{(11)} & \colhead{(12)} & \colhead{(13)} }
\startdata
\hline
\multicolumn{13}{c}{Density peaks detected by 5$\sigma$} \\
\hline
8+9 & 2:27:06 & -4:20:30 & 3.304 & 18 & 2.07 & 2.83 &	3.09 & 19.19 & 6.47 & 4830.9 &  3.140 & 28.12 \\
11 & 2:26.35 & -4.22:24 &	3.299 & 1 & 2.20 & 0.99 & 1.04 & 3.07 & 3.02  & 156.5 & 0.092 & 12.06 \\
13 & 2:26:20 &	-4.23:35 &	3.300 & 4 & 2.09 & 1.52 & 1.14 & 6.91 & 5.20  & 697.2 & 0.466 & 22.13 \\
14 & 2:26:34 &	-4.26:38 &	3.309 & 2 & 1.81 & 0.96 & 0.84 & 4.05 & 4.50  & 358.3 & 0.220 & 17.50 \\
\hline
\multicolumn{13}{c}{protostructure detected by 2$\sigma$} \\
\hline
PS & 2:26:48 & -4:20:56 & 3.303 & 45 & 0.92 & 8.71 & 6.43 & 28.61 & 3.76 & 52951.0& 26.082 & 10.83 \\
\enddata
\tablecomments{(1) are the ID of peaks; (2), (3) and (4) are the RA, Dec and z of the barycenter of peaks; (5) the number of spectroscopic members; (6) is the average $\delta_{gal, 3D}$ in 3D maps; (7)-(9) are the effective radius of RA, Dec and z ; (10) is the elongation; (11) \& (12) are total volume and total mass $M_{tot}$; (13) is the average $\delta_{gal}$ derived by correcting columns (6) by the elongation in column (10) of this table. We note that Peak [5] is not detected, and Peak [8] and [9] are blended using a 5$\sigma$ in 3D overdensity maps. }
\end{deluxetable*}

Besides the evolutionary path of the two RAGNs that are found in the protostructure, we are also interested in the protostructure itself. Thanks to new spectra observations from Keck/MOSFIRE that gives us a larger data set of spectroscopically-confirmed galaxies and the new high-fidelity galaxy density mapping technique, we found this protostructure contains several overdensity peaks connected by filaments, as shown in the 2D map in Figure \ref{fig:2dmap}. 
In these two sections, we perform a simple exercise to understand the evolutionary status of individual density peaks and the protostructure. 
To facilitate this exercise, we construct the 3D overdensity maps (``3D cube'' hereafter) to quantify the structure of each peak along the line-of-sight (l.o.s). 
Even using the 3D cube, the volume computed is still probably an overestimate, since it is artificially elongated along the l.o.s., due to the combined effect of the induced peculiar motions of the member galaxies, the depth of the redshift slices, and the photometric redshift error. Thus, we apply a correction to the volume of these structures for this elongation factor. 

The 3D cube is constructed across the redshift range of 3.24 $\leq$ z $\leq$ 3.38, with overlapping redshift slices with a depth of 7.9 pMpc, which corresponds to $\delta$z$\sim$0.037 at z$\sim3.3$, running in steps of $\delta$z = 0.005 (see details on this algorithm in Section \ref{sec:env}.). 
Following \citet{Cucciati2018}, we identify the density peaks and the protostructure in the 3D cube by considering only the regions of space with log(1+$\delta_\mathrm{gal}$) above 5$\sigma$ and 2$\sigma$, respectively. More details on the average galaxy density and $\sigma$ are present in Appendix \ref{app:3dmaps}. 
We computed the barycenter of each peak by weighting the (x, y, z) position of each pixel belonging to the peak by its $\delta_{gal}$. For each detection, we measure the volume by adding up the volume of all the contiguous pixels bounded by the 5$\sigma$/2$\sigma$ surface and compute the average overdensity $\langle\delta_{gal}\rangle$ of all pixels. 
The total mass of each detection is calculated following Eq. 1 in \citet{Cucciati2018} $M_{tot} = \rho_m V (1+\delta_m)$, where $\rho_\mathrm{m}$ is the matter density evaluated at z=0, V is the comoving volume of each detection, and $\delta_m$ is the matter overdensity. 
We compute $\delta_m$ by using the relation $\delta_m$ = $\langle\delta_{gal}\rangle/b$, adopting a bias factor b = 2.68 as derived in \cite{Durkalec2015} at z $\sim 3$ near the redshift covered in this paper. 

To take into account the artificial elongation along the l.o.s, as mentioned above, we adopt the same approach to correct for this elongation factor as \citet{Cucciati2018} under the assumption that on average our peaks should have roughly the same size in the l.o.s. dimension as the average of the two transverse dimensions. For each of the three dimensions, we measured an effective radius $R_e$ defined as $R_{e,x} = \sqrt{\Sigma_i w_i (x_i - x_{peak})^2 / \Sigma_i(w_i)}$) (and same for $R_{e, y}$ and $R_{e,z}$), where the sum is over all the pixels belonging to the given peak, the weight $w_i$ is the value of $\delta_{gal}$, $x_i$ the position in cMpc along the x-axis and $x_{peak}$ is the barycenter of the peak along the x-axis. 
We defined the elongation $E_{z/xy}$ for each peak as the ratio between $R_{e,z}$ and $R_{e,xy}$, where $R_{e,xy}$ is the mean between $R_{e,x}$ and $R_{e,y}$. The corrected volume is the measured volume divided by $E_{z/xy}$. Given that the elongation has the opposite and compensating effects of increasing
the volume and decreasing $\delta_{gal}$, the $M_{tot}$ remains the same. We further derive the $\langle\delta_{corr, gal}\rangle$ for each peak using the corrected volume and $M_{tot}$. 
All properties derived from 3D maps are reported in Table \ref{tab:3dpeaks}. We note that peak [5] is not detected in the 3D map, and peak [8] is blended with peak [9] within the surface bounded by a 5$\sigma$ detection threshold.
For more details of 3D cube and the 2$\sigma$/5$\sigma$ contours of each density peaks, see Appendix \ref{app:3dmaps}. 

\subsection{The predicted evolution of the protostructure and consistent peaks} 
\label{sec:evolution_protocluster}

To predict the evolutionary status of these peaks and how peaks with similar overdensities would evolve with time, we adopt the framework of the spherical collapse model following the recipe described in \citet{Cucciati2018}. 
According to the spherical collapse model, any spherical overdensity will evolve like a sub-universe, with a matter-energy density higher than the critical overdensity at any given epoch. 
Here, we assume the average matter overdensity $\langle \delta_m \rangle$ of our peaks are in a non-linear regime. To simplify the calculation, we compute the evolution of an overdensity in linear regime by transforming $\langle \delta_m \rangle$ into their corresponding values in linear regime $\langle \delta_{m, L}\rangle$ \citep{Padmanabhan1993}. The $\langle \delta_m \rangle$ is defined as $\langle \delta_{corr,gal}\rangle$/b, where $\langle \delta_{corr,gal}\rangle$ is the average $\delta_{gal}$ corrected by the elongation reported in Table \ref{tab:3dpeaks} and b is 2.68. 
In the spherical linear collapse model, the overdense sphere passes through three specific evolutionary steps. The first one is the point of turn-around ($\delta_{L,ta} \simeq 1.062$), when the overdense sphere stops expanding and starts collapsing, becoming a gravitationally bound structure. After the turn-around, when the radius of the sphere becomes half of the radius at turn-around, the overdense sphere reaches virialization at $\delta_{L, vir} \simeq 1.58$. The last step is the moment of maximum collapse, which theoretically happens when its radius becomes zero with an infinite density. In the real universe, the collapse stops when the system satisfies the virial theorem at $\delta_{L,c} \simeq 1.686$. Here, we are interested in the time/redshift of turn-around and collapse as $t_{ta}$/z$_{ta}$ and $t_c$/z$_c$, respectively.

\begin{figure}
	\centering
	\includegraphics[width=\columnwidth]{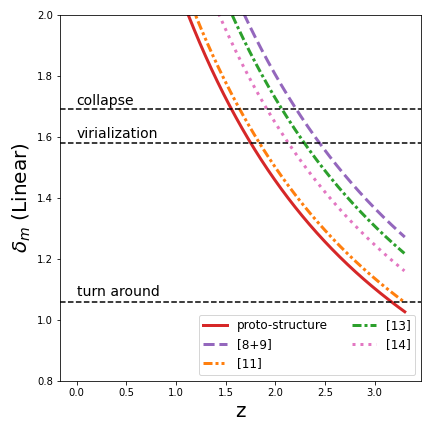}
	\caption{Evolution of $\delta_m$ for the density peaks and protostructure detected by 2$\sigma$/5$\sigma$ threshold in the 3D cube, with different line styles representing different peaks as in the legend. The evolution is computed in a linear regime for a $\Lambda$CDM Universe. For each peak, we start tracking the evolution from the barycenter redshift (column 4 in Table \ref{tab:3dpeaks}), and a starting $\langle\delta_m\rangle$ computed from $\langle\delta_{corr, gal}\rangle$ (column 13 in Table \ref{tab:3dpeaks}) and transformed into linear regime. The horizontal lines represent $\delta_{L,ta} \simeq 1.062$, $\delta_{L,vir} \simeq 1.58$ and $\delta_{L,c} \simeq 1.686$. Additional details are reported in Section \ref{sec:evolution_protocluster}.\label{fig:SCmodel}}
\end{figure}

The evolution of an overdense sphere at $z_2$ can be derived from a knowing $\delta_L(z_1)$ as: $$\delta_L(z_2) = \delta_L(z_1) \frac{D_+(z_2)}{D_+(z_1)} $$where $D_+(z)$ is the growing mode. In a $\Lambda$CDM universe, the linear growth factor g is defined as $g = D_+(z)/a$, where $a = (1 + z)^{-1}$ is the cosmic scale factor  (e.g. \citealp{Carroll1992, Hamilton2001}). 
For each peak, we start tracking the evolution from its barycenter redshift z$_{peak}$ and $\langle \delta_{m} \rangle$ (Table \ref{tab:evolution}). 
Figure \ref{fig:SCmodel} shows the evolution of the density peaks and the protostructure. 
Table 5 lists the values of z$_{ta}$ and z$_c$, together with the time elapsed from z$_{obs}$ to these two redshifts. 
It is not unexpected that the most evolved is peak [8+9], the original protocluster and the one housing the proto-BCG. According to the spherical collapse model, peak [8+9] will be a virialised system by z $\sim$ 2.2, that is, in 1.1 Gyr from the epoch of observation. The least evolved peak is [11] housing the radio-BCG. It is at the turn-around moment and will take another $\sim$ 2.1 Gyr to virialise. 
\citet{Cucciati2018} estimated the evolutionary status of seven density peaks within 2.4 $\lesssim$ z $\lesssim$ 2.5. These density peaks are found to be connected by filaments in a complex proto-supercluster ``Hyperion''. Under the same method/threshold, their peaks, on average, will take another $\Delta t_c=1.33\sim4.37$ Gyr to virialise. Our estimated time of collapse is slightly shorter, on average, than what \citet{Cucciati2018} estimated for their peaks. 
As for the protostructure, it has the lowest $\langle \delta_m \rangle$ than any density peak, which is not unexpected in such a large structure. It would take 0.1 Gy to reach a turn-around and then another $\sim$2.2 Gyr to virialise. These results indicate that the two peaks housing the RAGNs are in different evolutionary status and would evolve in parallel for another $\sim$0.1 Gyr before their collapsing process. 

Admittedly, our estimation is over-simplified, assuming that the peaks are isolated overdense spheres evolving in the absence of interactions with other density peaks.
In reality, the evolution will be more complex, due to possible merger events and accreting mass/subcomponents/galaxies during their lifetimes. 
Our exercise, presented here for heuristic purpose, is mainly to internally compare the evolutionary status of each peak and the protostructure associated with RAGNs.

\begin{deluxetable}{cc|ccccc}
\tablecolumns{7}
\tablewidth{0pt}
\tablecaption{Evolution of the density peaks and protostructure according to the spherical collapse model in the linear regime \label{tab:evolution}}
\tablehead{
\colhead{ID} & \colhead{z}& \colhead{$\delta_{m}$} & \colhead{z$_{ta}$} & \colhead{z$_{c}$} & \colhead{$\Delta t_{ta}$} & \colhead{$\Delta t_{c}$}  \vspace{-0.2cm}\\
 & &  & & & \colhead{[Gyr]} &  \colhead{[Gyr]}  \vspace{-0.2cm}\\
\colhead{(1)} & \colhead{(2)} & \colhead{(3)} & \colhead{(4)} & \colhead{(5)} & \colhead{(6)} & \colhead{(7)} }
\startdata
8+9 & 3.304 & 10.49 & $>$z$_{obs}$ & 2.21 &  - & 1.08 \\
11 & 3.299 & 4.50 & 3.296 & 1.64 & 0.002 & 2.07 \\
13 & 3.300 & 8.26 & $>$z$_{obs}$ & 2.07 & - & 1.29  \\
14 & 3.308 & 6.53 & $>$z$_{obs}$ & 1.92 & - & 1.53 \\
PS & 3.303 & 4.04 & 3.168 & 1.56 &  0.097 & 2.27  \\
\enddata
\tablecomments{\footnotesize Columns (1) and (2) are the ID and the barycenter redshift of the peak, as in Table \ref{tab:3dpeaks}. Column (3) is the average matter overdensity derived from the average galaxy overdensity of column (13) of Table \ref{tab:3dpeaks}. Columns (4) and (5) are the redshifts when the overdensity reaches the overdensity of turn-around and collapse, respectively. Columns (6) and (7) are the corresponding time intervals $t$ since the redshift of observation z$_{obs}$ (column 2) to the redshifts of turn-around and collapse. When $\mathrm{z}_{ta} < \mathrm{z}_{obs}$ the turn-around has already been reached before the redshift of observation, and in these cases the corresponding $t$ has not been computed. See Section \ref{sec:evolution_protocluster} for more details.}
\end{deluxetable}

\subsection{The formation and evolution of RAGNs}
\label{sec:evolution_RAGN}

\begin{figure*} 
	\centering
	\includegraphics[width=\textwidth]{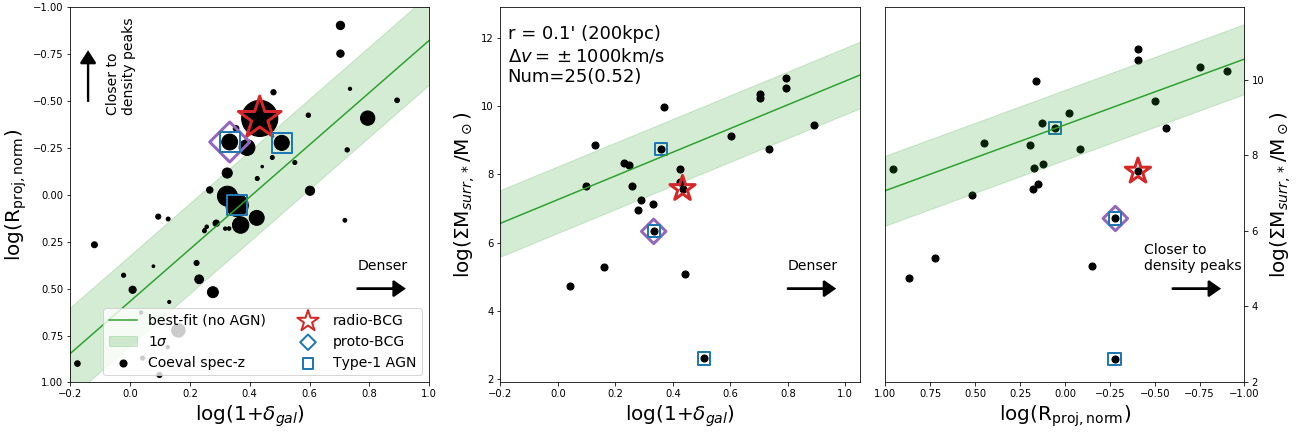}
	\caption{\textit{Left}: The log($1+\delta_\mathrm{gal}$) versus log(R$_\mathrm{proj, norm}$) for coeval spec-$z$ galaxies, with radio-BCG, proto-BCG and Type-1 AGNs marked. The size of dots are scaled by their stellar mass. The best-fitted line of coeval spec-$z$ galaxies, excluding RAGNs/Type-1 AGNs, is shown in green solid lines with $1\sigma$ envelopes as shaded regions. \textit{Middle} \& \textit{Right}: The log($1+\delta_\mathrm{gal}$) (\textit{middle}) and R$_\mathrm{proj, norm}$ (\textit{right}) as a function of the sum of stellar mass of neighboring galaxies for coeval spec-$z$ galaxies with neighboring galaxies in the selected cylindrical volume. The best-fitted lines of coeval spec-$z$ galaxies with neighboring galaxies, excluding RAGNs/Type-1 AGNs, are shown in green solid lines with $1\sigma$ envelopes as shaded regions. Note that 48\% of coeval spec-$z$ galaxies that do not have neighboring galaxies in the selected cylindrical volume are not shown in these figures. \label{fig:sumSM}}
\end{figure*}

In Section \ref{sec:env}, we find that the RAGNs/Type-1 AGNs reside in moderate local densities but fairly close to the density peaks. However, we would expect a higher local density at distances closer to a density peak. 
To visualize this, we show log($1+\delta_\mathrm{gal}$) versus log(R$_\mathrm{proj, norm}$) for coeval spec-$z$ galaxies in the left panel of Figure \ref{fig:sumSM}. The size of dots is scaled by their stellar mass. The best-fitted line of coeval spec-$z$ galaxies, excluding RAGNs/Type-1 AGNs, is shown as a green solid line with the $1\sigma$ envelope as the shaded region. As expected, the RAGNs are offset from the best-fitted line at the higher end of the $1\sigma$ envelop. This confirms that RAGNs have lower local densities than is to be expected from their locations with respect to their parent peak centers.  

Another result from Section \ref{sec:env} is that the RAGNs/Type-1 AGNs have higher stellar masses than coeval spec-$z$ galaxies found at the same moderate local densities and those fairly close to the density peaks. 
This evidence could suggest that merging events have happened in these RAGNs/Type-1 AGNs, which have served to boost their stellar mass and lower their local density at the same time retaining their global environment. In addition, mergers are sometimes thought to be responsible for transporting available gas on host galaxy scales to the central regions, triggering activity in the nucleus (e.g., \citealp{Ellison2011, Chiaberge2015, Tadhunter2016, Padovani2016}, though see, e.g., \citealp{Kocevski2012, Shah2020} for an alternative view).

In order to test if merging events could plausibly explain the observed results, we compare the amount of surrounding stellar mass of RAGNs/Type-1 AGNs to that of the coeval spec-$z$ galaxies. If a merging event has happened without continuous galaxy infall from the outskirts, it would increase the host stellar mass and decrease the amount of surrounding stellar mass. In detail, for each coeval spec-$z$ galaxy, we select its neighboring spec-$z$ galaxies/photo-$z$ objects in a cylindrical volume centered on it having r$_\mathrm{proj}$ = 0.1$\arcmin$ ($\sim$200 pkpc at z=3.3) and a fixed depth of $\Delta v = \pm$ 1000 km/s. The amount of surrounding stellar mass is calculated as the sum of the stellar mass of its neighboring spec-$z$ galaxies and photo-$z$ objects, as log($\Sigma$$M_\mathrm{surr, *}$/$M_\odot$). To account for the uncertainty of z$_\mathrm{phot}$ for photo-$z$ objects, their stellar masses are weighted by the probability of the photo-$z$ objects being in the cylindrical volume. The probability is estimated by integrating the p(z) of z$_\mathrm{phot}$ estimated in the SED fitting on the $\Delta v$ interval. Due to the incompleteness of spec-$z$ and photo-$z$ samples, we restrict the spec-$z$ galaxies/photo-$z$ objects to those with stellar mass $\geq10^9$ $M_\odot$ (see Section \ref{sec:spec}). The log($\Sigma$$M_\mathrm{surr, *}$/$M_\odot$) of coeval spec-$z$ galaxies is shown as functions of log(1+$\delta_\mathrm{gal}$) and log(R$_\mathrm{proj, norm}$) in the right two panels of Figure \ref{fig:sumSM}. We note that the 48\% of coeval spec-$z$ galaxies that do not have neighboring galaxies in the selected cylindrical volume are not shown in these figures. We adopt the same fitting approach as in Section \ref{sec:env} to the log($\Sigma$$M_\mathrm{surr, *}$/$M_\odot$) versus log(1+$\delta_\mathrm{gal}$)/log(R$_\mathrm{proj, norm}$) of coeval spec-$z$ galaxies that have neighboring galaxies, excluding RAGNs/Type-1 AGNs. The best-fitted lines are shown as green lines with the 1$\sigma$ envelopes as shaded regions. 
We find that, at the same $\delta_\mathrm{gal}$ and R$_\mathrm{proj, norm}$, RAGNs have smaller amounts of surrounding stellar mass, relative to those coeval spec-$z$ galaxies with neighboring galaxies. 
This result agrees with our hypothesis that merger-induced stellar mass growth in the host would decrease the amount of surrounding stellar mass compared to galaxies in the similar environments in this protostructure. 

To consider whether these results are biased by more massive galaxies or galaxies residing in denser regions, we re-run the analysis in Section \ref{sec:env} using only these coeval spec-$z$ galaxies with neighboring galaxies. These coeval spec-$z$ galaxies are spread over the full coeval spec-$z$ sample in the local/global density - stellar mass diagrams, slightly skewing to higher stellar masses and denser regions. The proto-/radio-BCG are still offset from these best-fitted lines with slightly smaller significance level of 1.8$\sigma$/3.3$\sigma$ and 1.7$\sigma$/3.2$\sigma$ in $\delta_\mathrm{gal}$ and R$_\mathrm{proj, norm}$, respectively. These results alleviate the concern of the representativeness of those galaxies with neighboring galaxies in the local/global density - stellar mass phase diagrams. 
As for the other two Type-1 AGNs, they show different results. One of them has an average amount of surrounding stellar mass comparable to the general trend of coeval spec-$z$ galaxies in both local and global environments. The other Type-1 AGN has a strong deficit. Due to such diverse results, we do not make any conclusions on them. 

Admittedly, this is a very simple test and based on a small sample. The 200 pkpc radius is a large scale for the merging scenario. Although larger values are apt to wash out any real signal, a smaller radius would limit the number of neighboring galaxies. If we choose a smaller radius (i.e., 100 pkpc), only 10 coeval spec-$z$ galaxies have neighboring galaxies, and none of the AGNs would have any neighboring galaxies. 
In addition, the difference in the masses of the two RAGNs relative to other coeval spec-$z$ galaxies is in excess of their amount of stellar material that is predicted to have surrounded them. This indicates that the merging scenario might not be able to fully explain the stellar mass gained in the RAGN hosts with respect to expectations. Other mechanisms, such as a bursty or sustained star-formation at a high level (e.g., \citealp{Forrest2020}) and/or additional galaxies infall from the outskirts at much earlier times, may be needed. 

Unfortunately, we do not have definitive evidence to support our merging scenario. Observations, imaging, such as an Atacama Large Millimetre Array (ALMA) map taken in a higher-resolution configuration, \emph{Hubble Space Telescope}, or \emph{James Webb Space Telescope} imaging could be used to confirm or deny the proposed scenario.

Beside the merging scenario, in the local universe, Bondi accretion, due to the interaction between galaxies and hot ICM, is thought to be the dominant process igniting and sustaining RAGN activity. \citep{Bondi1952, Allen2006, Balmaverde2008, Fujita2016}. These low-z RAGNs are hosted by massive galaxies and reside in clusters. They seem to share similar properties to RAGNs in this paper, for at least the Radio-BCG. 
The hot X-ray emitting intracluster medium (ICM) has been observed in (proto-)clusters at 2 $\la$ z $\la$ 3 (e.g. \citealp{Gobat2011, Wang2016, Valentino2016}). 
The pervasive AGN activity observed in many high-redshift protoclusters suggests a possible mechanism to pre-heat the proto-ICM (e.g., \citealp{Hilton2012, Kravtsov2012} and references therein). However, we can not draw any conclusion on this scenario for what concerns the peaks of this protostructure, since no ICM has been detected at z $>$ 3. 

\subsection{The detections of protoclusters around RAGNs} \label{sec:PCdetection}

Powerful radio galaxies have been extensively used for high-redshift cluster searches (e.g., \citealp{Rigby2014, Koyama2014, Overzier2008, Matsuda2009, Hatch2011b, Hayashi2012, Cooke2014, Shimakawa2014, Hatch2011a, Galametz2012, Wylezalek2013}.
Among these studies, the Clusters Around Radio-Loud AGN (CARLA) survey is the largest statistical study, designed to investigate the environments of powerful RAGNs. They studied 419 very powerful RAGNs lying at 1.3 $<$ z $<$ 3.2 and having a 500MHz luminosity $> 10^{27.5} W Hz^{-1}$. By comparing the surface density of IRAC-selected sources in the RAGNs and blank fields, they found 55\% (10\%) of the RAGN fields are overdense at the 2$\sigma$(5$\sigma$) level \citep{Wylezalek2013, Wylezalek2014}. 
Furthermore, \citet{Hatch2014} compared the RAGN sample in CARLA with radio-quiet galaxies matched in mass and redshift. They found the environments of RAGNs are significantly denser, which suggested that the dense Mpc-scale environment fosters the formation of a radio-jet from an AGN. 

While there have been many studies on high-redshift radio galaxies and their environments, few of them are targeted at lower power radio sources. 
This work is the first time that an RAGN at L$_{1.4GHz} \sim 10^{25}$ W Hz$^{-1}$ has been discovered in a high-z protocluster, a posteriori. 
This is quite promising news for future large area radio surveys, which will reach very faint flux densities \citep{Padovani2016}. For detection limit down to $\sim$mJy, we could detect radio sources similar to radio-BCG that have radio luminosity of L$_\mathrm{1.4GHz} \sim 10^{26} \mathrm{W Hz}^{-1}$ at z $\sim$ 3. Following the same approach as this work, we expect to detect overdense regions around these low-luminosity radio sources, though less massive than protocluster, like proto-groups having $M_{tot}$ similar to peak [11] at $\sim1\times10^{13}$ $M_\odot$. 
In addition, such studies can be motivated by lower redshift studies. Low-luminosity RAGNs also tend to reside in rich groups and clusters with L$_\mathrm{1.4GHz} \sim 10^{24}$ W Hz$^{-1}$ at z $\sim$ 1 \citep{Shen2017}. 
\citet{Castignani2014} searched for galaxy clusters in z $\sim$ 1-2 using low power radio galaxies L$_\mathrm{1.4GHz} \sim 10^{25.3}$ W Hz$^{-1}$ at z $\sim $1.1 as beacons. They found $\sim$ 70\% radio galaxies reside in overdensities, independent of radio luminosity. 
At slightly higher redshift, \citet{Daddi2017} use the deep VLA 3GHz imaging with $S_\mathrm{3GHz} >$ 8$\mu$Jy in the COSMOS field, cross matched to spectroscopically-confirmed cluster members of a protocluster Cl J1001 at z$_{spec}$ = 2.506 \citep{Wang2016}. They detected six radio galaxies within 10$\arcsec$ radius from the center of protocluster, corresponding to $\sim$300 pkpc at z=2.5. These studies all suggest that large samples of high-redshift proto-clusters/groups could be found when deeper radio surveys, such as LOFAR and SKA, are available.

\section{Summary} \label{sec:sum}

Using previously reported observations from the VVDS and VUDS surveys, new spectral observations from Keck/MOSFIRE, and new radio observations from JVLA, we report here on the discovery of two RAGNs in a massive and complex protostructure in the region surrounding the PCl J0227-0421 protocluster at z=3.29. We interpret the properties of the host and environmental properties of RAGNs, as follows:

 \begin{itemize}
    	\item One RAGN, termed ``proto-BCG'', is hosted by the brightest galaxy in the optical/NIR, was previously found to be extremely blue in restframe color, very massive in stellar mass (10$^{10.9}$ $M_\odot$), and detected in Herschel to have an obscured SFR of $\sim700 M_\odot/yr$. It is also a Type-1 AGN. These properties indicate strong star-formation activity in the proto-BCG with co-evolution of the AGN and star-formation. 
	
	\item A second, newly discovered RAGN host, termed ``radio-BCG'', is hosted by the reddest and most massive galaxy (10$^{11.62}$ $M_\odot$) in the sample of coeval spec-$z$ galaxies. The properties of the host indicate that it is near quiescence and is very likely to be the dominant progenitor of what will be the z$\sim$0 BCG. 

	\item We find that the two RAGNs live in fairly average local environments relative to the coeval spec-$z$ galaxies in and around the protostructure. However, they are within the effective radius of their parent-peaks and fairly close to the center of their parent-peaks compared to the coeval spec-$z$ galaxies. They are more massive than galaxies that would be found in similar local and global environment in this protostructureprotostructure. We propose a scenario where merging might already have happened in both cases that has lowered the local density of their surrounding area and boosted their stellar mass. We perform a test to assess the plausibility of this scenario by calculating the amount of surrounding stellar mass. We find a compelling result that merger-induced stellar mass growth in the host would decrease the amount of surrounding stellar mass compared to galaxies in the similar environments in this protostructure. 
	
	\item We estimate the total mass of individual peaks, $M_{tot}$, based on their average galaxy density, and found a range of masses from $\sim0.1\times 10^{14} M_\odot$ to $\sim3.1\times 10^{14} M_\odot$. The estimated $M_{tot}$ of the protostructure is $\sim2.6\times 10^{15} M_\odot$. If we assume that the peaks are going to evolve separately, without accretion/merger events, the spherical collapse model predicts that these peaks have already started or are about to start their collapse phase, and they will be all virialised by redshift z $\sim$ 1.6. 
	
\end{itemize}

This work is the first time that two RAGNs at low-luminosity (L$_{1.4GHz} \sim 10^{25}$ W Hz$^{-1}$) have been found and studied within a high redshift protostructure. For future large area deep radio surveys, we would expect to detect a large sample of overdense regions, proto-clusters/groups, around lower-luminosity radio sources.

\section*{Acknowledgements}
We thank the anonymous referee for the valuable comments and Yongquan Xue for the informative discussion. LS and GL acknowledge the grant from the National Key R\&D Program of China (2016YFA0400702), the National Natural Science Foundation of China (No. 11673020 and No. 11421303). LS also acknowledge the NSFC Grants No. 12003030. WF acknowledge the NSFC Grants No. 11773024. 
This material is based upon work supported by the National Science Foundation under Grant No. 1411943. 
This study is based on data taken with the Karl G. Jansky Very Large Array which is operated by the National Radio Astronomy Observatory. The National Radio Astronomy Observatory is a facility of the National Science Foundation operated under cooperative agreement by Associated Universities, Inc. 
This work is based on data products made available at the CESAM data center, Laboratoire d'Astrophysique de Marseille. This work partly uses observations obtained with MegaPrime/MegaCam, a joint project of CFHT and CEA/DAPNIA, at the Canada-France-Hawaii Telescope (CFHT) , which is operated by the National Research Council (NRC) of Canada, the Institut National des Sciences de l'Univers of the Centre National de la Recherche Scientifique (CNRS) of France, and the University of Hawaii. This work is based in part on data products produced at TERAPIX and the Canadian Astronomy Data Centre as part of the Canada-France-Hawaii Telescope Legacy Survey, a collaborative project of NRC and CNRS. 
The spectrographic data presented herein were obtained at the W.M. Keck Observatory, which is operated as a scientific partnership among the California Institute of Technology, the University of California, and the National Aeronautics and Space Administration. The Observatory was made possible by the generous financial support of the W.M. Keck Foundation. 
We wish to thank the indigenous Hawaiian community for allowing us to be guests on their sacred mountain, a privilege, without with, this work would not have been possible. We are most fortunate to be able to conduct observations from this site.

\appendix

\section{Details in 3D overdensity maps}\label{app:3dmaps}

\begin{figure*}
	\centering
	\includegraphics[width=\textwidth]{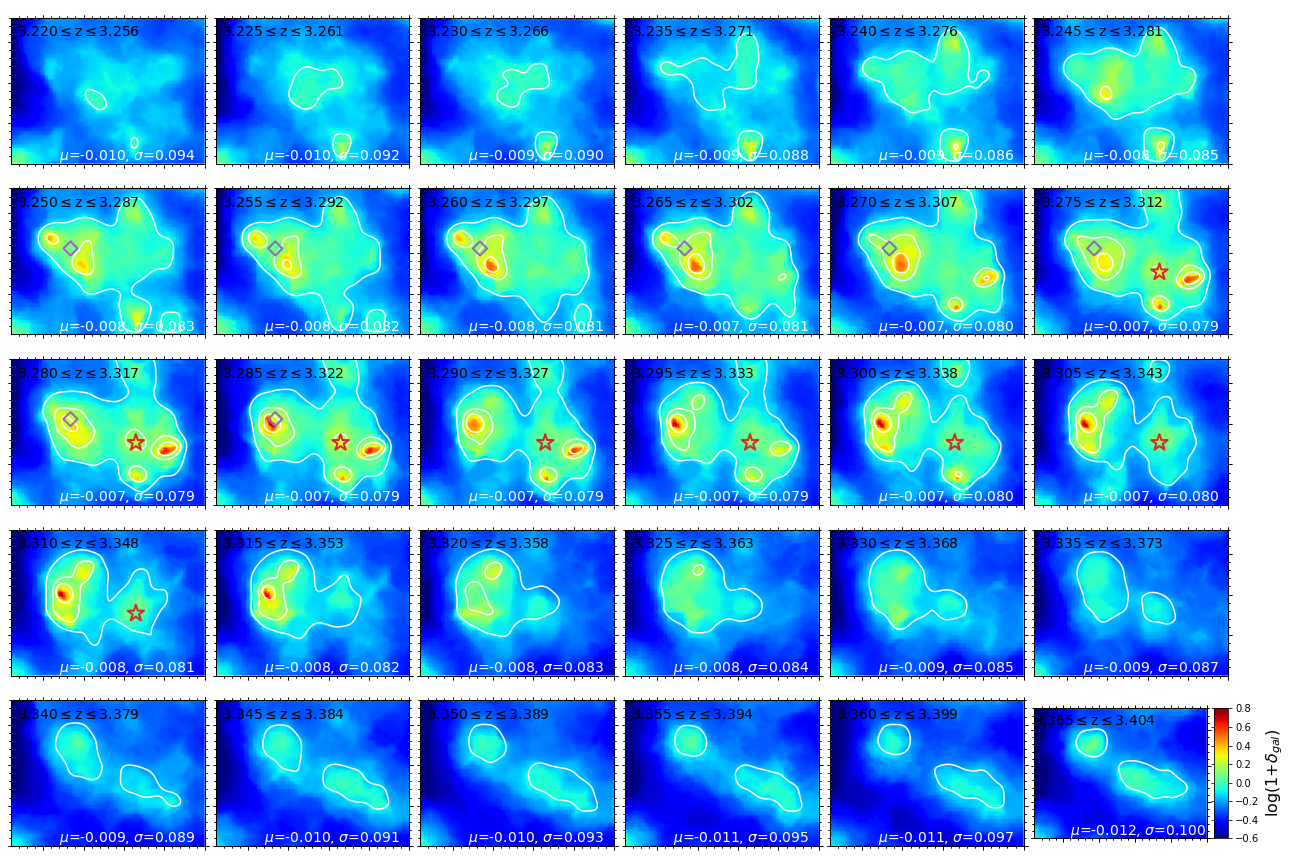}
	\caption{3D smoothed overdensity maps with color scaled by log(1+$\delta_\mathrm{gal}$). The color code is the same as Figure \ref{fig:2dmap}. White contours are 1, 2, 3$\sigma$ contours calculated with respect to each redshift slice. The two RAGNs are marked in open colored markers. \label{fig:3dmaps}}
\end{figure*}

\begin{figure*}
         \centering
         \includegraphics[width=0.65\textwidth]{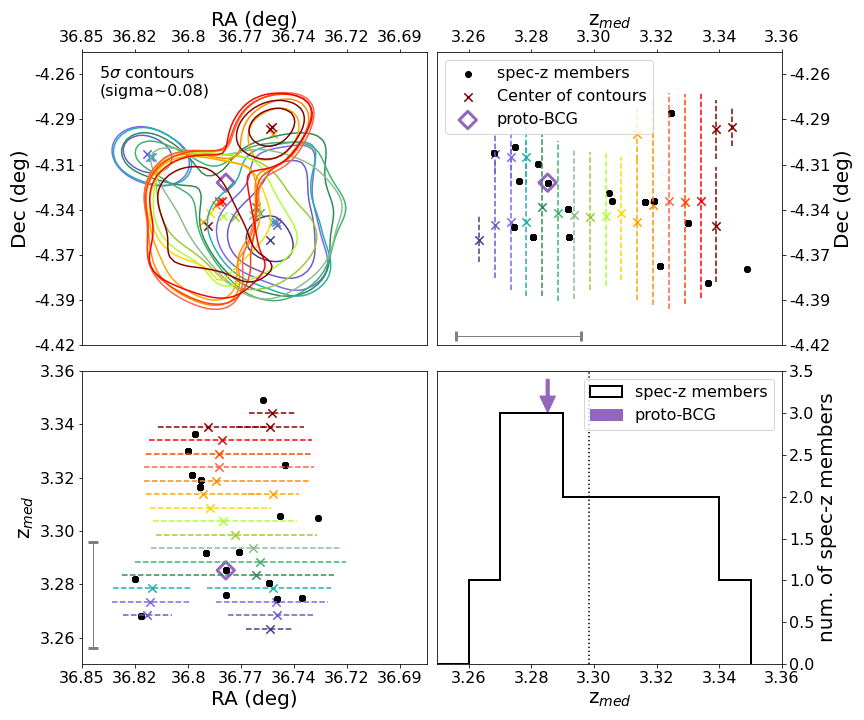}
          \includegraphics[width=0.65\textwidth]{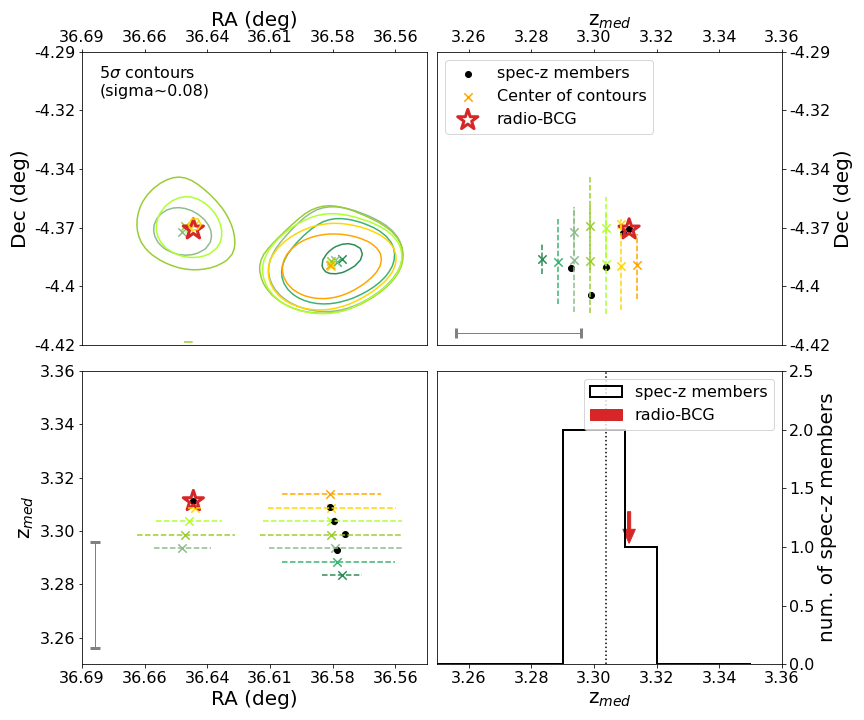}
	 \caption{For the main peak (a) and the radio-BCG peak (b), the \textit{top-left} panels show the projection on the RA-Dec plane of the 1$\sigma$ contours; the different colors indicate the different redshift slices (from blue to red, they go from the lowest to the highest redshift). Colored crosses are the RA-DEC barycenter of the contours, with the same color code as the contours. Open purple diamond (a) and red star (b) are the position of radio-BCG and proto-BCG, respectively.  The \textit{top-right} and \textit{bottom-left} panels are projections on the z-Dec and RA-z plane of the same contours shown in the top-left panel, with the same color code. The grey dots are spectroscopically-confirmed galaxies within the 1$\sigma$ contours. The colored crosses and open markers are as in the top-left panel. The redshift spanning range of a galaxy is indicated by vertical/horizontal errorbars. The black histogram in the \textit{Bottom-right} panels represents the redshift distribution of the spectroscopic galaxies which fall in the 1$\sigma$ contour. The vertical dotted line indicates the center along the l.o.s (the top x-axis is the same as the one in the top-right panel), and the RAGNs are indicated by purple and red arrows. \label{fig:3dcontours}}
\end{figure*}

We show here the 3D cube for the protostructure. As we mentioned in Section \ref{sec:env}, we construct 3-dimensional (3D) overdensity maps in overlapping redshift slices with a depth of 7.9 pMpc, which corresponds to $\delta$z$\sim$0.037 at z$\sim$3.3, running in steps of $\delta$z=0.005. In Figure \ref{fig:3dmaps}, we show VMC slices from z = 3.22 to z = 3.40, overlaid by 2, 5, 7$\sigma$ contours above the median overdensity $\mu$ for each redshift slice. The redshift range, $\mu$ and $\sigma$ are shown in each slice (see more details on the calculation of $\mu$ and $\sigma$ in \citealp{Cucciati2018}). The RAGNs are marked in the associated redshift slice.

In Figure \ref{fig:3dcontours}, we show the projections on the RA-Dec, z-Dec and RA-z planes of the two density peaks that are housing the two RAGNs to highlight their complex shape. The projections that we show include the 5$\sigma$ contours in each redshift slice colored by their redshift. The center of each contour is plotted by a colored cross. The position of the coeval spec-$z$ galaxies within these 5$\sigma$ contours and proto-BCG/radio-BCG are plotted in grey dots and open purple/red signs. Since the redshift slices overlap, a galaxy is actually included in multiple redshift slices. We indicate the redshift spanning range of a galaxy in the z-Dec/RA-z plots by vertical/horizontal errorbars. Redshift histograms of coeval spec-$z$ members are shown in the lower right panel with their median z$_\mathrm{spec}$ shown in dotted vertical lines and redshift of proto-BCG/radio-BCG indicated by purple/red arrows. 

\section{Spectrum of the proto-BCG and the SED fitting of RAGN hosts}\label{app:sed}

\begin{figure}
	\centering
	\includegraphics[width=\columnwidth]{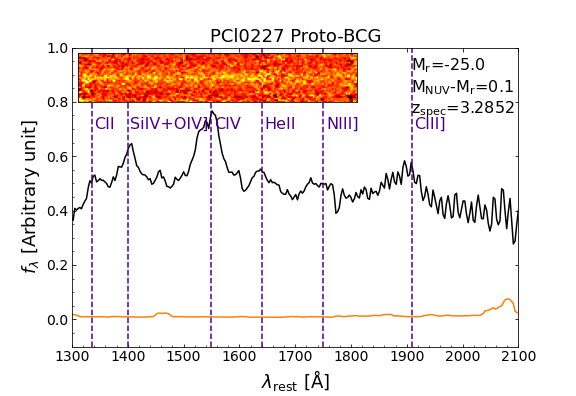}
	\caption{The 2d/1d spectrum of the proto-BCG. The spectrum is obtained by VLT/VIMOS taken as part of VVDS. The black line is the one-dimensional flux density spectrum and the orange line is the formal uncertainty spectrum. Important spectral features are marked. The SED best-fitted results for this galaxy are shown on the top right corner. The top inset panel shows the two-dimensional spectrum. This spectrum contains several high-ionization emission features with their FWHMs $\ge$ 1000 km s$^{-1}$, which indicates the proto-BCG is clearly a Type 1 AGN. Note the 1d spectrum of the proto-BCG has been presented in \citet{Lemaux2014a} (the left third panel of their Figure 6, ID=20465339).  \label{fig:spec_protoBCG}}
\end{figure}

\begin{figure*}
	\centering
	\includegraphics[width=0.8\textwidth]{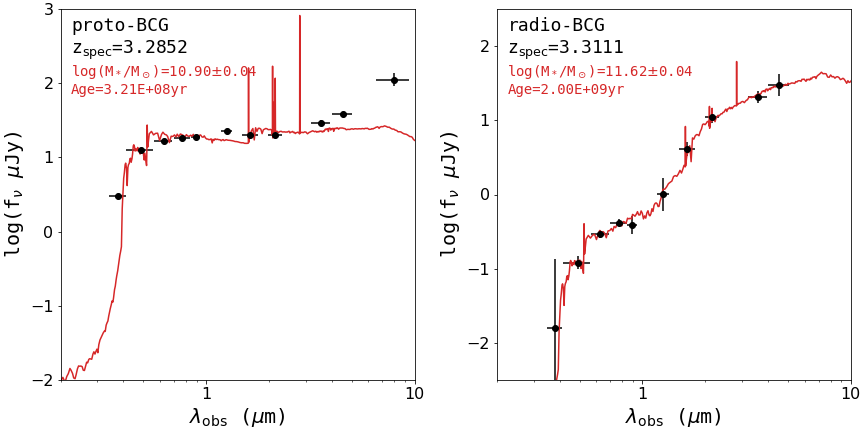}
	\caption{Observed-frame optical/NIR  broadband photometry of the proto-BCG (\textit{left}) and radio-BCG (\textit{right}) with their best-fit galaxy SED fitting overplotted in red. The error bars in the x-direction indicate the wavelength extent of the various filters. The z$_\mathrm{spec}$, best-fit stellar mass, and luminosity-weighted stellar age are shown in the upper left corner of each plot. Although the luminosity-weighted age from this fitting is a highly uncertain quantity (see, e.g., \citealp{Thomas2017}), the large differences in the ages recovered for the two RAGNs are well beyond the typical error and highlight the different evolutionary stage of the two galaxies. The offset shown in the near-infrared part of the proto-BCG SED fits is due to the effect of the AGN. If those are excised in the SED fitting, all physical parameters are statistically consistent with the values reported above. Note that the SED of the proto-BCG has been presented in \citet{Lemaux2014a} (their Figure 11). See \citet{Lemaux2014a} Appendix A for more details on the SED fitting process. \label{fig:sed}}
\end{figure*}

We present the 2d/1d spectra of the proto-BCG in Figure \ref{fig:spec_protoBCG}, for comparison with the 2d/1d spectra of radio-BCG in Figure \ref{fig:spec}.  To show the extremely different properties of the host galaxies of the two RAGNs, we show the observed photometry along with the SED fits of the two RAGN hosts in Figure \ref{fig:sed}.

\begin{thebibliography}{}
	\expandafter\ifx\csname natexlab\endcsname\relax\def\natexlab#1{#1}\fi
	\providecommand{\url}[1]{\href{#1}{#1}}
	\providecommand{\dodoi}[1]{doi:~\href{http://doi.org/#1}{\nolinkurl{#1}}}
	\providecommand{\doeprint}[1]{\href{http://ascl.net/#1}{\nolinkurl{http://ascl.net/#1}}}
	\providecommand{\doarXiv}[1]{\href{https://arxiv.org/abs/#1}{\nolinkurl{https://arxiv.org/abs/#1}}}
	
	\bibitem[{{Allen} {et~al.}(2006){Allen}, {Dunn}, {Fabian}, {Taylor}, \&
	  {Reynolds}}]{Allen2006}
	{Allen}, S.~W., {Dunn}, R.~J.~H., {Fabian}, A.~C., {Taylor}, G.~B., \&
	  {Reynolds}, C.~S. 2006, \mnras, 372, 21,
	  \dodoi{10.1111/j.1365-2966.2006.10778.x}
	
	\bibitem[{{Arnouts} {et~al.}(1999){Arnouts}, {Cristiani}, {Moscardini},
	  {Matarrese}, {Lucchin}, {Fontana}, \& {Giallongo}}]{Arnouts1999}
	{Arnouts}, S., {Cristiani}, S., {Moscardini}, L., {et~al.} 1999, \mnras, 310,
	  540, \dodoi{10.1046/j.1365-8711.1999.02978.x}
	
	\bibitem[{{Ascaso} {et~al.}(2014){Ascaso}, {Lemaux}, {Lubin}, {Gal},
	  {Kocevski}, {Rumbaugh}, \& {Squires}}]{Ascaso2014}
	{Ascaso}, B., {Lemaux}, B.~C., {Lubin}, L.~M., {et~al.} 2014, \mnras, 442, 589,
	  \dodoi{10.1093/mnras/stu877}
	
	\bibitem[{{Balmaverde} {et~al.}(2008){Balmaverde}, {Baldi}, \&
	  {Capetti}}]{Balmaverde2008}
	{Balmaverde}, B., {Baldi}, R.~D., \& {Capetti}, A. 2008, \aap, 486, 119,
	  \dodoi{10.1051/0004-6361:200809810}
	
	\bibitem[{{Bell} {et~al.}(2003){Bell}, {McIntosh}, {Katz}, \&
	  {Weinberg}}]{Bell2003}
	{Bell}, E.~F., {McIntosh}, D.~H., {Katz}, N., \& {Weinberg}, M.~D. 2003, \apjs,
	  149, 289, \dodoi{10.1086/378847}
	
	\bibitem[{{Bertin} \& {Arnouts}(1996)}]{Bertin1996}
	{Bertin}, E., \& {Arnouts}, S. 1996, \aaps, 117, 393,
	  \dodoi{10.1051/aas:1996164}
	
	\bibitem[{{Best}(2004)}]{Best2004}
	{Best}, P.~N. 2004, \mnras, 351, 70, \dodoi{10.1111/j.1365-2966.2004.07752.x}
	
	\bibitem[{{Best} {et~al.}(2005){Best}, {Kauffmann}, {Heckman}, {Brinchmann},
	  {Charlot}, {Ivezi{\'c}}, \& {White}}]{Best2005}
	{Best}, P.~N., {Kauffmann}, G., {Heckman}, T.~M., {et~al.} 2005, \mnras, 362,
	  25, \dodoi{10.1111/j.1365-2966.2005.09192.x}
	
	\bibitem[{{Bielby} {et~al.}(2012){Bielby}, {Hudelot}, {McCracken}, {Ilbert},
	  {Daddi}, {Le F{\`e}vre}, {Gonzalez-Perez}, {Kneib}, {Marmo}, {Mellier},
	  {Salvato}, {Sanders}, \& {Willott}}]{Bielby2012}
	{Bielby}, R., {Hudelot}, P., {McCracken}, H.~J., {et~al.} 2012, \aap, 545, A23,
	  \dodoi{10.1051/0004-6361/201118547}
	
	\bibitem[{{Bleem} {et~al.}(2015){Bleem}, {Stalder}, {de Haan}, {Aird}, {Allen},
	  {Applegate}, {Ashby}, {Bautz}, {Bayliss}, {Benson}, {Bocquet}, {Brodwin},
	  {Carlstrom}, {Chang}, {Chiu}, {Cho}, {Clocchiatti}, {Crawford}, {Crites},
	  {Desai}, {Dietrich}, {Dobbs}, {Foley}, {Forman}, {George}, {Gladders},
	  {Gonzalez}, {Halverson}, {Hennig}, {Hoekstra}, {Holder}, {Holzapfel},
	  {Hrubes}, {Jones}, {Keisler}, {Knox}, {Lee}, {Leitch}, {Liu}, {Lueker},
	  {Luong-Van}, {Mantz}, {Marrone}, {McDonald}, {McMahon}, {Meyer}, {Mocanu},
	  {Mohr}, {Murray}, {Padin}, {Pryke}, {Reichardt}, {Rest}, {Ruel}, {Ruhl},
	  {Saliwanchik}, {Saro}, {Sayre}, {Schaffer}, {Schrabback}, {Shirokoff},
	  {Song}, {Spieler}, {Stanford}, {Staniszewski}, {Stark}, {Story}, {Stubbs},
	  {Vand erlinde}, {Vieira}, {Vikhlinin}, {Williamson}, {Zahn}, \&
	  {Zenteno}}]{Bleem2015}
	{Bleem}, L.~E., {Stalder}, B., {de Haan}, T., {et~al.} 2015, \apjs, 216, 27,
	  \dodoi{10.1088/0067-0049/216/2/27}
	
	\bibitem[{{Bondi}(1952)}]{Bondi1952}
	{Bondi}, H. 1952, \mnras, 112, 195, \dodoi{10.1093/mnras/112.2.195}
	
	\bibitem[{{Bondi} {et~al.}(2007){Bondi}, {Ciliegi}, {Venturi}, {Dallacasa},
	  {Bardelli}, {Zucca}, {Athreya}, {Gregorini}, {Zanichelli}, {Le F{\`e}vre},
	  {Contini}, {Garilli}, {Iovino}, {Temporin}, \& {Vergani}}]{Bondi2007}
	{Bondi}, M., {Ciliegi}, P., {Venturi}, T., {et~al.} 2007, \aap, 463, 519,
	  \dodoi{10.1051/0004-6361:20066428}
	
	\bibitem[{{Bonzini} {et~al.}(2015){Bonzini}, {Mainieri}, {Padovani},
	  {Andreani}, {Berta}, {Bethermin}, {Lutz}, {Rodighiero}, {Rosario}, {Tozzi},
	  \& {Vattakunnel}}]{Bonzini2015}
	{Bonzini}, M., {Mainieri}, V., {Padovani}, P., {et~al.} 2015, \mnras, 453,
	  1079, \dodoi{10.1093/mnras/stv1675}
	
	\bibitem[{{Boulade} {et~al.}(2003){Boulade}, {Charlot}, {Abbon}, {Aune},
	  {Borgeaud}, {Carton}, {Carty}, {Da Costa}, {Deschamps}, {Desforge},
	  {Eppell{\'e}}, {Gallais}, {Gosset}, {Granelli}, {Gros}, {de Kat}, {Loiseau},
	  {Ritou}, {Rouss{\'e}}, {Starzynski}, {Vignal}, \& {Vigroux}}]{Boulade2003}
	{Boulade}, O., {Charlot}, X., {Abbon}, P., {et~al.} 2003, in Society of
	  Photo-Optical Instrumentation Engineers (SPIE) Conference Series, Vol. 4841,
	  Instrument Design and Performance for Optical/Infrared Ground-based
	  Telescopes, ed. M.~{Iye} \& A.~F.~M. {Moorwood}, 72--81,
	  \dodoi{10.1117/12.459890}
	
	\bibitem[{{Bouwens} {et~al.}(2015){Bouwens}, {Illingworth}, {Oesch}, {Trenti},
	  {Labb{\'e}}, {Bradley}, {Carollo}, {van Dokkum}, {Gonzalez}, {Holwerda},
	  {Franx}, {Spitler}, {Smit}, \& {Magee}}]{Bouwens2015}
	{Bouwens}, R.~J., {Illingworth}, G.~D., {Oesch}, P.~A., {et~al.} 2015, \apj,
	  803, 34, \dodoi{10.1088/0004-637X/803/1/34}
	
	\bibitem[{{Carroll} {et~al.}(1992){Carroll}, {Press}, \&
	  {Turner}}]{Carroll1992}
	{Carroll}, S.~M., {Press}, W.~H., \& {Turner}, E.~L. 1992, \araa, 30, 499,
	  \dodoi{10.1146/annurev.aa.30.090192.002435}
	
	\bibitem[{{Castignani} {et~al.}(2014){Castignani}, {Chiaberge}, {Celotti},
	  {Norman}, \& {De Zotti}}]{Castignani2014}
	{Castignani}, G., {Chiaberge}, M., {Celotti}, A., {Norman}, C., \& {De Zotti},
	  G. 2014, \apj, 792, 114, \dodoi{10.1088/0004-637X/792/2/114}
	
	\bibitem[{{Chen} {et~al.}(2018){Chen}, {Brandt}, {Luo}, {Ranalli}, {Yang},
	  {Alexander}, {Bauer}, {Kelson}, {Lacy}, {Nyland}, {Tozzi}, {Vito},
	  {Cirasuolo}, {Gilli}, {Jarvis}, {Lehmer}, {Paolillo}, {Schneider}, {Shemmer},
	  {Smail}, {Sun}, {Tanaka}, {Vaccari}, {Vignali}, {Xue}, {Banerji}, {Chow},
	  {H{\"a}u{\ss}ler}, {Norris}, {Silverman}, \& {Trump}}]{Chen2018}
	{Chen}, C. T.~J., {Brandt}, W.~N., {Luo}, B., {et~al.} 2018, \mnras, 478, 2132,
	  \dodoi{10.1093/mnras/sty1036}
	
	\bibitem[{{Chiaberge} {et~al.}(2015){Chiaberge}, {Gilli}, {Lotz}, \&
	  {Norman}}]{Chiaberge2015}
	{Chiaberge}, M., {Gilli}, R., {Lotz}, J.~M., \& {Norman}, C. 2015, \apj, 806,
	  147, \dodoi{10.1088/0004-637X/806/2/147}
	
	\bibitem[{{Condon}(1992)}]{Condon1992}
	{Condon}, J.~J. 1992, \araa, 30, 575,
	  \dodoi{10.1146/annurev.aa.30.090192.003043}
	
	\bibitem[{{Cooke} {et~al.}(2014){Cooke}, {Hatch}, {Muldrew}, {Rigby}, \&
	  {Kurk}}]{Cooke2014}
	{Cooke}, E.~A., {Hatch}, N.~A., {Muldrew}, S.~I., {Rigby}, E.~E., \& {Kurk},
	  J.~D. 2014, \mnras, 440, 3262, \dodoi{10.1093/mnras/stu522}
	
	\bibitem[{{Cooke} {et~al.}(2015){Cooke}, {Hatch}, {Rettura}, {Wylezalek},
	  {Galametz}, {Stern}, {Brodwin}, {Muldrew}, {Almaini}, {Conselice},
	  {Eisenhardt}, {Hartley}, {Jarvis}, {Seymour}, \& {Stanford}}]{Cooke2015}
	{Cooke}, E.~A., {Hatch}, N.~A., {Rettura}, A., {et~al.} 2015, \mnras, 452,
	  2318, \dodoi{10.1093/mnras/stv1413}
	
	\bibitem[{{Cooke} {et~al.}(2016){Cooke}, {O'Dea}, {Baum}, {Tremblay}, {Cox}, \&
	  {Gladders}}]{Cooke2016}
	{Cooke}, K.~C., {O'Dea}, C.~P., {Baum}, S.~A., {et~al.} 2016, \apj, 833, 224,
	  \dodoi{10.3847/1538-4357/833/2/224}
	
	\bibitem[{{Cooper} {et~al.}(2007){Cooper}, {Newman}, {Coil}, {Croton}, {Gerke},
	  {Yan}, {Davis}, {Faber}, {Guhathakurta}, {Koo}, {Weiner}, \&
	  {Willmer}}]{Cooper2007}
	{Cooper}, M.~C., {Newman}, J.~A., {Coil}, A.~L., {et~al.} 2007, \mnras, 376,
	  1445, \dodoi{10.1111/j.1365-2966.2007.11534.x}
	
	\bibitem[{{Cucciati} {et~al.}(2014){Cucciati}, {Zamorani}, {Lemaux},
	  {Bardelli}, {Cimatti}, {Le F{\`e}vre}, {Cassata}, {Garilli}, {Le Brun},
	  {Maccagni}, {Pentericci}, {Tasca}, {Thomas}, {Vanzella}, {Zucca}, {Amorin},
	  {Capak}, {Cassar{\`a}}, {Castellano}, {Cuby}, {de la Torre}, {Durkalec},
	  {Fontana}, {Giavalisco}, {Grazian}, {Hathi}, {Ilbert}, {Moreau}, {Paltani},
	  {Ribeiro}, {Salvato}, {Schaerer}, {Scodeggio}, {Sommariva}, {Talia},
	  {Taniguchi}, {Tresse}, {Vergani}, {Wang}, {Charlot}, {Contini}, {Fotopoulou},
	  {L{\'o}pez-Sanjuan}, {Mellier}, \& {Scoville}}]{Cucciati2014}
	{Cucciati}, O., {Zamorani}, G., {Lemaux}, B.~C., {et~al.} 2014, \aap, 570, A16,
	  \dodoi{10.1051/0004-6361/201423811}
	
	\bibitem[{{Cucciati} {et~al.}(2017){Cucciati}, {Davidzon}, {Bolzonella},
	  {Granett}, {De Lucia}, {Branchini}, {Zamorani}, {Iovino}, {Garilli}, {Guzzo},
	  {Scodeggio}, {de la Torre}, {Abbas}, {Adami}, {Arnouts}, {Bottini}, {Cappi},
	  {Franzetti}, {Fritz}, {Krywult}, {Le Brun}, {Le F{\`e}vre}, {Maccagni},
	  {Ma{\l}ek}, {Marulli}, {Moutard}, {Polletta}, {Pollo}, {Tasca}, {Tojeiro},
	  {Vergani}, {Zanichelli}, {Bel}, {Blaizot}, {Coupon}, {Hawken}, {Ilbert},
	  {Moscardini}, {Peacock}, \& {Gargiulo}}]{Cucciati2017}
	{Cucciati}, O., {Davidzon}, I., {Bolzonella}, M., {et~al.} 2017, \aap, 602,
	  A15, \dodoi{10.1051/0004-6361/201630113}
	
	\bibitem[{{Cucciati} {et~al.}(2018){Cucciati}, {Lemaux}, {Zamorani}, {Le
	  F{\`e}vre}, {Tasca}, {Hathi}, {Lee}, {Bardelli}, {Cassata}, {Garilli}, {Le
	  Brun}, {Maccagni}, {Pentericci}, {Thomas}, {Vanzella}, {Zucca}, {Lubin},
	  {Amorin}, {Cassar{\`a}}, {Cimatti}, {Talia}, {Vergani}, {Koekemoer}, {Pforr},
	  \& {Salvato}}]{Cucciati2018}
	{Cucciati}, O., {Lemaux}, B.~C., {Zamorani}, G., {et~al.} 2018, \aap, 619, A49,
	  \dodoi{10.1051/0004-6361/201833655}
	
	\bibitem[{{Daddi} {et~al.}(2017){Daddi}, {Jin}, {Strazzullo}, {Sargent},
	  {Wang}, {Ferrari}, {Schinnerer}, {Smol{\v{c}}i{\'c}}, {Calabr{\'o}},
	  {Coogan}, {Delhaize}, {Delvecchio}, {Elbaz}, {Gobat}, {Gu}, {Liu}, {Novak},
	  \& {Valentino}}]{Daddi2017}
	{Daddi}, E., {Jin}, S., {Strazzullo}, V., {et~al.} 2017, \apjl, 846, L31,
	  \dodoi{10.3847/2041-8213/aa8808}
	
	\bibitem[{{Darvish} {et~al.}(2020){Darvish}, {Scoville}, {Martin}, {Sobral},
	  {Mobasher}, {Rettura}, {Matthee}, {Capak}, {Chartab}, {Hemmati}, {Masters},
	  {Nayyeri}, {O'Sullivan}, {Paulino-Afonso}, {Sattari}, {Shahidi}, {Salvato},
	  {Lemaux}, {F{\`e}vre}, \& {Cucciati}}]{Darvish2020}
	{Darvish}, B., {Scoville}, N.~Z., {Martin}, C., {et~al.} 2020, \apj, 892, 8,
	  \dodoi{10.3847/1538-4357/ab75c3}
	
	\bibitem[{{Davis} {et~al.}(2003){Davis}, {Faber}, {Newman}, {Phillips},
	  {Ellis}, {Steidel}, {Conselice}, {Coil}, {Finkbeiner}, {Koo}, {Guhathakurta},
	  {Weiner}, {Schiavon}, {Willmer}, {Kaiser}, {Luppino}, {Wirth}, {Connolly},
	  {Eisenhardt}, {Cooper}, \& {Gerke}}]{Davis2003}
	{Davis}, M., {Faber}, S.~M., {Newman}, J., {et~al.} 2003, in \procspie, Vol.
	  4834, Discoveries and Research Prospects from 6- to 10-Meter-Class Telescopes
	  II, ed. P.~{Guhathakurta}, 161--172, \dodoi{10.1117/12.457897}
	
	\bibitem[{{Durkalec} {et~al.}(2015){Durkalec}, {Le F{\`e}vre}, {Pollo}, {de la
	  Torre}, {Cassata}, {Garilli}, {Le Brun}, {Lemaux}, {Maccagni}, {Pentericci},
	  {Tasca}, {Thomas}, {Vanzella}, {Zamorani}, {Zucca}, {Amor{\'\i}n},
	  {Bardelli}, {Cassar{\`a}}, {Castellano}, {Cimatti}, {Cucciati}, {Fontana},
	  {Giavalisco}, {Grazian}, {Hathi}, {Ilbert}, {Paltani}, {Ribeiro}, {Schaerer},
	  {Scodeggio}, {Sommariva}, {Talia}, {Tresse}, {Vergani}, {Capak}, {Charlot},
	  {Contini}, {Cuby}, {Dunlop}, {Fotopoulou}, {Koekemoer}, {L{\'o}pez-Sanjuan},
	  {Mellier}, {Pforr}, {Salvato}, {Scoville}, {Taniguchi}, \&
	  {Wang}}]{Durkalec2015}
	{Durkalec}, A., {Le F{\`e}vre}, O., {Pollo}, A., {et~al.} 2015, \aap, 583,
	  A128, \dodoi{10.1051/0004-6361/201425343}
	
	\bibitem[{{Ellison} {et~al.}(2011){Ellison}, {Patton}, {Mendel}, \&
	  {Scudder}}]{Ellison2011}
	{Ellison}, S.~L., {Patton}, D.~R., {Mendel}, J.~T., \& {Scudder}, J.~M. 2011,
	  \mnras, 418, 2043, \dodoi{10.1111/j.1365-2966.2011.19624.x}
	
	\bibitem[{{Everett} {et~al.}(2020){Everett}, {Zhang}, {Crawford}, {Vieira},
	  {Aravena}, {Archipley}, {Austermann}, {Benson}, {Bleem}, {Carlstrom},
	  {Chang}, {Chapman}, {Crites}, {de Haan}, {Dobbs}, {George}, {Halverson},
	  {Harrington}, {Holder}, {Holzapfel}, {Hrubes}, {Knox}, {Lee}, {Luong-Van},
	  {Mangian}, {Marrone}, {McMahon}, {Meyer}, {Mocanu}, {Mohr}, {Natoli},
	  {Padin}, {Pryke}, {Reichardt}, {Reuter}, {Ruhl}, {Sayre}, {Schaffer},
	  {Shirokoff}, {Spilker}, {Stalder}, {Staniszewski}, {Stark}, {Story},
	  {Switzer}, {Vanderlinde}, {Wei{\ss}}, \& {Williamson}}]{Everett2020}
	{Everett}, W.~B., {Zhang}, L., {Crawford}, T.~M., {et~al.} 2020, \apj, 900, 55,
	  \dodoi{10.3847/1538-4357/ab9df7}
	
	\bibitem[{{Faber} {et~al.}(2003){Faber}, {Phillips}, {Kibrick}, {Alcott},
	  {Allen}, {Burrous}, {Cantrall}, {Clarke}, {Coil}, {Cowley}, {Davis}, {Deich},
	  {Dietsch}, {Gilmore}, {Harper}, {Hilyard}, {Lewis}, {McVeigh}, {Newman},
	  {Osborne}, {Schiavon}, {Stover}, {Tucker}, {Wallace}, {Wei}, {Wirth}, \&
	  {Wright}}]{Faber2003}
	{Faber}, S.~M., {Phillips}, A.~C., {Kibrick}, R.~I., {et~al.} 2003, in
	  \procspie, Vol. 4841, Instrument Design and Performance for Optical/Infrared
	  Ground-based Telescopes, ed. M.~{Iye} \& A.~F.~M. {Moorwood}, 1657--1669,
	  \dodoi{10.1117/12.460346}
	
	\bibitem[{{Fazio} {et~al.}(2004){Fazio}, {Hora}, {Allen}, {Ashby}, {Barmby},
	  {Deutsch}, {Huang}, {Kleiner}, {Marengo}, {Megeath}, {Melnick}, {Pahre},
	  {Patten}, {Polizotti}, {Smith}, {Taylor}, {Wang}, {Willner}, {Hoffmann},
	  {Pipher}, {Forrest}, {McMurty}, {McCreight}, {McKelvey}, {McMurray}, {Koch},
	  {Moseley}, {Arendt}, {Mentzell}, {Marx}, {Losch}, {Mayman}, {Eichhorn},
	  {Krebs}, {Jhabvala}, {Gezari}, {Fixsen}, {Flores}, {Shakoorzadeh}, {Jungo},
	  {Hakun}, {Workman}, {Karpati}, {Kichak}, {Whitley}, {Mann}, {Tollestrup},
	  {Eisenhardt}, {Stern}, {Gorjian}, {Bhattacharya}, {Carey}, {Nelson},
	  {Glaccum}, {Lacy}, {Lowrance}, {Laine}, {Reach}, {Stauffer}, {Surace},
	  {Wilson}, {Wright}, {Hoffman}, {Domingo}, \& {Cohen}}]{Fazio2004}
	{Fazio}, G.~G., {Hora}, J.~L., {Allen}, L.~E., {et~al.} 2004, \apjs, 154, 10,
	  \dodoi{10.1086/422843}
	
	\bibitem[{{Forrest} {et~al.}(2020){Forrest}, {Marsan}, {Annunziatella},
	  {Wilson}, {Muzzin}, {Marchesini}, {Cooper}, {Chan}, {McConachie}, {Gomez},
	  {Kado-Fong}, {Barbera}, {Lange-Vagle}, {Nantais}, {Nonino}, {Saracco},
	  {Stefanon}, \& {van der Burg}}]{Forrest2020}
	{Forrest}, B., {Marsan}, Z.~C., {Annunziatella}, M., {et~al.} 2020, \apj, 903,
	  47, \dodoi{10.3847/1538-4357/abb819}
	
	\bibitem[{{Fujita} {et~al.}(2016){Fujita}, {Kawakatu}, \&
	  {Shlosman}}]{Fujita2016}
	{Fujita}, Y., {Kawakatu}, N., \& {Shlosman}, I. 2016, \pasj, 68, 26,
	  \dodoi{10.1093/pasj/psw012}
	
	\bibitem[{{Fukugita} {et~al.}(1996){Fukugita}, {Ichikawa}, {Gunn}, {Doi},
	  {Shimasaku}, \& {Schneider}}]{Fukugita1996}
	{Fukugita}, M., {Ichikawa}, T., {Gunn}, J.~E., {et~al.} 1996, \aj, 111, 1748,
	  \dodoi{10.1086/117915}
	
	\bibitem[{{Fuller} {et~al.}(2020){Fuller}, {Lemaux}, {Brada{\v{c}}}, {Hoag},
	  {Schmidt}, {Huang}, {Strait}, {Mason}, {Treu}, {Pentericci}, {Trenti},
	  {Henry}, \& {Malkan}}]{Fuller2020}
	{Fuller}, S., {Lemaux}, B.~C., {Brada{\v{c}}}, M., {et~al.} 2020, \apj, 896,
	  156, \dodoi{10.3847/1538-4357/ab959f}
	
	\bibitem[{{Galametz} {et~al.}(2012){Galametz}, {Stern}, {De Breuck}, {Hatch},
	  {Mayo}, {Miley}, {Rettura}, {Seymour}, {Stanford}, \&
	  {Vernet}}]{Galametz2012}
	{Galametz}, A., {Stern}, D., {De Breuck}, C., {et~al.} 2012, \apj, 749, 169,
	  \dodoi{10.1088/0004-637X/749/2/169}
	
	\bibitem[{{Gobat} {et~al.}(2011){Gobat}, {Daddi}, {Onodera}, {Finoguenov},
	  {Renzini}, {Arimoto}, {Bouwens}, {Brusa}, {Chary}, {Cimatti}, {Dickinson},
	  {Kong}, \& {Mignoli}}]{Gobat2011}
	{Gobat}, R., {Daddi}, E., {Onodera}, M., {et~al.} 2011, \aap, 526, A133,
	  \dodoi{10.1051/0004-6361/201016084}
	
	\bibitem[{{Griffin} {et~al.}(2010){Griffin}, {Abergel}, {Abreu}, {Ade},
	  {Andr{\'e}}, {Augueres}, {Babbedge}, {Bae}, {Baillie}, {Baluteau}, {Barlow},
	  {Bendo}, {Benielli}, {Bock}, {Bonhomme}, {Brisbin}, {Brockley-Blatt},
	  {Caldwell}, {Cara}, {Castro-Rodriguez}, {Cerulli}, {Chanial}, {Chen},
	  {Clark}, {Clements}, {Clerc}, {Coker}, {Communal}, {Conversi}, {Cox},
	  {Crumb}, {Cunningham}, {Daly}, {Davis}, {de Antoni}, {Delderfield}, {Devin},
	  {di Giorgio}, {Didschuns}, {Dohlen}, {Donati}, {Dowell}, {Dowell}, {Duband},
	  {Dumaye}, {Emery}, {Ferlet}, {Ferrand}, {Fontignie}, {Fox}, {Franceschini},
	  {Frerking}, {Fulton}, {Garcia}, {Gastaud}, {Gear}, {Glenn}, {Goizel},
	  {Griffin}, {Grundy}, {Guest}, {Guillemet}, {Hargrave}, {Harwit}, {Hastings},
	  {Hatziminaoglou}, {Herman}, {Hinde}, {Hristov}, {Huang}, {Imhof}, {Isaak},
	  {Israelsson}, {Ivison}, {Jennings}, {Kiernan}, {King}, {Lange}, {Latter},
	  {Laurent}, {Laurent}, {Leeks}, {Lellouch}, {Levenson}, {Li}, {Li},
	  {Lilienthal}, {Lim}, {Liu}, {Lu}, {Madden}, {Mainetti}, {Marliani}, {McKay},
	  {Mercier}, {Molinari}, {Morris}, {Moseley}, {Mulder}, {Mur}, {Naylor},
	  {Nguyen}, {O'Halloran}, {Oliver}, {Olofsson}, {Olofsson}, {Orfei}, {Page},
	  {Pain}, {Panuzzo}, {Papageorgiou}, {Parks}, {Parr-Burman}, {Pearce},
	  {Pearson}, {P{\'e}rez-Fournon}, {Pinsard}, {Pisano}, {Podosek}, {Pohlen},
	  {Polehampton}, {Pouliquen}, {Rigopoulou}, {Rizzo}, {Roseboom}, {Roussel},
	  {Rowan-Robinson}, {Rownd}, {Saraceno}, {Sauvage}, {Savage}, {Savini},
	  {Sawyer}, {Scharmberg}, {Schmitt}, {Schneider}, {Schulz}, {Schwartz},
	  {Shafer}, {Shupe}, {Sibthorpe}, {Sidher}, {Smith}, {Smith}, {Smith},
	  {Spencer}, {Stobie}, {Sudiwala}, {Sukhatme}, {Surace}, {Stevens}, {Swinyard},
	  {Trichas}, {Tourette}, {Triou}, {Tseng}, {Tucker}, {Turner}, {Vaccari},
	  {Valtchanov}, {Vigroux}, {Virique}, {Voellmer}, {Walker}, {Ward}, {Waskett},
	  {Weilert}, {Wesson}, {White}, {Whitehouse}, {Wilson}, {Winter}, {Woodcraft},
	  {Wright}, {Xu}, {Zavagno}, {Zemcov}, {Zhang}, \& {Zonca}}]{Griffin2010}
	{Griffin}, M.~J., {Abergel}, A., {Abreu}, A., {et~al.} 2010, \aap, 518, L3,
	  \dodoi{10.1051/0004-6361/201014519}
	
	\bibitem[{{Hamilton}(2001)}]{Hamilton2001}
	{Hamilton}, A.~J.~S. 2001, \mnras, 322, 419,
	  \dodoi{10.1046/j.1365-8711.2001.04137.x}
	
	\bibitem[{{Hatch} {et~al.}(2011{\natexlab{a}}){Hatch}, {Kurk}, {Pentericci},
	  {Venemans}, {Kuiper}, {Miley}, \& {R{\"o}ttgering}}]{Hatch2011b}
	{Hatch}, N.~A., {Kurk}, J.~D., {Pentericci}, L., {et~al.} 2011{\natexlab{a}},
	  \mnras, 415, 2993, \dodoi{10.1111/j.1365-2966.2011.18735.x}
	
	\bibitem[{{Hatch} {et~al.}(2011{\natexlab{b}}){Hatch}, {De Breuck}, {Galametz},
	  {Miley}, {Overzier}, {R{\"o}ttgering}, {Doherty}, {Kodama}, {Kurk},
	  {Seymour}, {Venemans}, {Vernet}, \& {Zirm}}]{Hatch2011a}
	{Hatch}, N.~A., {De Breuck}, C., {Galametz}, A., {et~al.} 2011{\natexlab{b}},
	  \mnras, 410, 1537, \dodoi{10.1111/j.1365-2966.2010.17538.x}
	
	\bibitem[{{Hatch} {et~al.}(2014){Hatch}, {Wylezalek}, {Kurk}, {Stern}, {De
	  Breuck}, {Jarvis}, {Galametz}, {Gonzalez}, {Hartley}, {Mortlock}, {Seymour},
	  \& {Stevens}}]{Hatch2014}
	{Hatch}, N.~A., {Wylezalek}, D., {Kurk}, J.~D., {et~al.} 2014, \mnras, 445,
	  280, \dodoi{10.1093/mnras/stu1725}
	
	\bibitem[{{Hayashi} {et~al.}(2012){Hayashi}, {Kodama}, {Tadaki}, {Koyama}, \&
	  {Tanaka}}]{Hayashi2012}
	{Hayashi}, M., {Kodama}, T., {Tadaki}, K.-i., {Koyama}, Y., \& {Tanaka}, I.
	  2012, \apj, 757, 15, \dodoi{10.1088/0004-637X/757/1/15}
	
	\bibitem[{{Higuchi} {et~al.}(2019){Higuchi}, {Ouchi}, {Ono}, {Shibuya},
	  {Toshikawa}, {Harikane}, {Kojima}, {Chiang}, {Egami}, {Kashikawa},
	  {Overzier}, {Konno}, {Inoue}, {Hasegawa}, {Fujimoto}, {Goto}, {Ishikawa},
	  {Ito}, {Komiyama}, \& {Tanaka}}]{Higuchi2019}
	{Higuchi}, R., {Ouchi}, M., {Ono}, Y., {et~al.} 2019, \apj, 879, 28,
	  \dodoi{10.3847/1538-4357/ab2192}
	
	\bibitem[{{Hilton} {et~al.}(2012){Hilton}, {Romer}, {Kay}, {Mehrtens},
	  {Lloyd-Davies}, {Thomas}, {Short}, {Mayers}, {Rooney}, {Stott}, {Collins},
	  {Harrison}, {Hoyle}, {Liddle}, {Mann}, {Miller}, {Sahl{\'e}n}, {Viana},
	  {Davidson}, {Hosmer}, {Nichol}, {Sabirli}, {Stanford}, \&
	  {West}}]{Hilton2012}
	{Hilton}, M., {Romer}, A.~K., {Kay}, S.~T., {et~al.} 2012, \mnras, 424, 2086,
	  \dodoi{10.1111/j.1365-2966.2012.21359.x}
	
	\bibitem[{{Hovatta} {et~al.}(2014){Hovatta}, {Aller}, {Aller}, {Clausen-Brown},
	  {Homan}, {Kovalev}, {Lister}, {Pushkarev}, \& {Savolainen}}]{Hovatta2014}
	{Hovatta}, T., {Aller}, M.~F., {Aller}, H.~D., {et~al.} 2014, \aj, 147, 143,
	  \dodoi{10.1088/0004-6256/147/6/143}
	
	\bibitem[{{Hu} {et~al.}(2021){Hu}, {Wang}, {Infante}, {Rhoads}, {Zheng},
	  {Yang}, {Malhotra}, {Barrientos}, {Jiang}, {Gonz{\'a}lez-L{\'o}pez},
	  {Prieto}, {Perez}, {Hibon}, {Galaz}, {Coughlin}, {Harish}, {Kong}, {Kang},
	  {Khostovan}, {Pharo}, {Valdes}, {Wold}, {Walker}, \& {Zheng}}]{Hu2021}
	{Hu}, W., {Wang}, J., {Infante}, L., {et~al.} 2021, Nature Astronomy,
	  \dodoi{10.1038/s41550-020-01291-y}
	
	\bibitem[{{Hung} {et~al.}(2020){Hung}, {Lemaux}, {Gal}, {Tomczak}, {Lubin},
	  {Cucciati}, {Pelliccia}, {Shen}, {Le F{\`e}vre}, {Wu}, {Kocevski}, {Mei}, \&
	  {Squires}}]{Hung2020}
	{Hung}, D., {Lemaux}, B.~C., {Gal}, R.~R., {et~al.} 2020, \mnras, 491, 5524,
	  \dodoi{10.1093/mnras/stz3164}
	
	\bibitem[{{Ilbert} {et~al.}(2006){Ilbert}, {Arnouts}, {McCracken},
	  {Bolzonella}, {Bertin}, {Le F{\`e}vre}, {Mellier}, {Zamorani}, {Pell{\`o}},
	  {Iovino}, {Tresse}, {Le Brun}, {Bottini}, {Garilli}, {Maccagni}, {Picat},
	  {Scaramella}, {Scodeggio}, {Vettolani}, {Zanichelli}, {Adami}, {Bardelli},
	  {Cappi}, {Charlot}, {Ciliegi}, {Contini}, {Cucciati}, {Foucaud}, {Franzetti},
	  {Gavignaud}, {Guzzo}, {Marano}, {Marinoni}, {Mazure}, {Meneux}, {Merighi},
	  {Paltani}, {Pollo}, {Pozzetti}, {Radovich}, {Zucca}, {Bondi}, {Bongiorno},
	  {Busarello}, {de La Torre}, {Gregorini}, {Lamareille}, {Mathez}, {Merluzzi},
	  {Ripepi}, {Rizzo}, \& {Vergani}}]{Ilbert2006}
	{Ilbert}, O., {Arnouts}, S., {McCracken}, H.~J., {et~al.} 2006, Astronomy and
	  Astrophysics, 457, 841, \dodoi{10.1051/0004-6361:20065138}
	
	\bibitem[{{Ilbert} {et~al.}(2009){Ilbert}, {Capak}, {Salvato}, {Aussel},
	  {McCracken}, {Sanders}, {Scoville}, {Kartaltepe}, {Arnouts}, {Le Floc'h},
	  {Mobasher}, {Taniguchi}, {Lamareille}, {Leauthaud}, {Sasaki}, {Thompson},
	  {Zamojski}, {Zamorani}, {Bardelli}, {Bolzonella}, {Bongiorno}, {Brusa},
	  {Caputi}, {Carollo}, {Contini}, {Cook}, {Coppa}, {Cucciati}, {de la Torre},
	  {de Ravel}, {Franzetti}, {Garilli}, {Hasinger}, {Iovino}, {Kampczyk},
	  {Kneib}, {Knobel}, {Kovac}, {Le Borgne}, {Le Brun}, {Le F{\`e}vre}, {Lilly},
	  {Looper}, {Maier}, {Mainieri}, {Mellier}, {Mignoli}, {Murayama}, {Pell{\`o}},
	  {Peng}, {P{\'e}rez-Montero}, {Renzini}, {Ricciardelli}, {Schiminovich},
	  {Scodeggio}, {Shioya}, {Silverman}, {Surace}, {Tanaka}, {Tasca}, {Tresse},
	  {Vergani}, \& {Zucca}}]{Ilbert2009}
	{Ilbert}, O., {Capak}, P., {Salvato}, M., {et~al.} 2009, \apj, 690, 1236,
	  \dodoi{10.1088/0004-637X/690/2/1236}
	
	\bibitem[{{Ito} {et~al.}(2019){Ito}, {Kashikawa}, {Toshikawa}, {Overzier},
	  {Tanaka}, {Kubo}, {Shibuya}, {Ishikawa}, {Onoue}, {Uchiyama}, {Liang},
	  {Higuchi}, {Martin}, {Lee}, {Komiyama}, \& {Huang}}]{Ito2019}
	{Ito}, K., {Kashikawa}, N., {Toshikawa}, J., {et~al.} 2019, \apj, 878, 68,
	  \dodoi{10.3847/1538-4357/ab1f0c}
	
	\bibitem[{{Kauffmann} {et~al.}(2008){Kauffmann}, {Heckman}, \&
	  {Best}}]{Kauffmann2008}
	{Kauffmann}, G., {Heckman}, T.~M., \& {Best}, P.~N. 2008, \mnras, 384, 953,
	  \dodoi{10.1111/j.1365-2966.2007.12752.x}
	
	\bibitem[{{Kauffmann} {et~al.}(2004){Kauffmann}, {White}, {Heckman},
	  {M{\'e}nard}, {Brinchmann}, {Charlot}, {Tremonti}, \&
	  {Brinkmann}}]{Kauffmann2004}
	{Kauffmann}, G., {White}, S.~D.~M., {Heckman}, T.~M., {et~al.} 2004, \mnras,
	  353, 713, \dodoi{10.1111/j.1365-2966.2004.08117.x}
	
	\bibitem[{{Kocevski} {et~al.}(2012){Kocevski}, {Faber}, {Mozena}, {Koekemoer},
	  {Nandra}, {Rangel}, {Laird}, {Brusa}, {Wuyts}, {Trump}, {Koo}, {Somerville},
	  {Bell}, {Lotz}, {Alexander}, {Bournaud}, {Conselice}, {Dahlen}, {Dekel},
	  {Donley}, {Dunlop}, {Finoguenov}, {Georgakakis}, {Giavalisco}, {Guo},
	  {Grogin}, {Hathi}, {Juneau}, {Kartaltepe}, {Lucas}, {McGrath}, {McIntosh},
	  {Mobasher}, {Robaina}, {Rosario}, {Straughn}, {van der Wel}, \&
	  {Villforth}}]{Kocevski2012}
	{Kocevski}, D.~D., {Faber}, S.~M., {Mozena}, M., {et~al.} 2012, \apj, 744, 148,
	  \dodoi{10.1088/0004-637X/744/2/148}
	
	\bibitem[{{Koyama} {et~al.}(2014){Koyama}, {Kodama}, {Tadaki}, {Hayashi},
	  {Tanaka}, \& {Shimakawa}}]{Koyama2014}
	{Koyama}, Y., {Kodama}, T., {Tadaki}, K.-i., {et~al.} 2014, \apj, 789, 18,
	  \dodoi{10.1088/0004-637X/789/1/18}
	
	\bibitem[{{Kravtsov} \& {Borgani}(2012)}]{Kravtsov2012}
	{Kravtsov}, A.~V., \& {Borgani}, S. 2012, \araa, 50, 353,
	  \dodoi{10.1146/annurev-astro-081811-125502}
	
	\bibitem[{{Le F{\`e}vre} {et~al.}(2003){Le F{\`e}vre}, {Saisse}, {Mancini},
	  {Brau-Nogue}, {Caputi}, {Castinel}, {D'Odorico}, {Garilli}, {Kissler-Patig},
	  {Lucuix}, {Mancini}, {Pauget}, {Sciarretta}, {Scodeggio}, {Tresse}, \&
	  {Vettolani}}]{LeFevre2003}
	{Le F{\`e}vre}, O., {Saisse}, M., {Mancini}, D., {et~al.} 2003, in Society of
	  Photo-Optical Instrumentation Engineers (SPIE) Conference Series, Vol. 4841,
	  \procspie, ed. M.~{Iye} \& A.~F.~M. {Moorwood}, 1670--1681,
	  \dodoi{10.1117/12.460959}
	
	\bibitem[{{Le F{\`e}vre} {et~al.}(2004){Le F{\`e}vre}, {Vettolani}, {Paltani},
	  {Tresse}, {Zamorani}, {Le Brun}, {Moreau}, {Bottini}, {Maccagni}, {Picat},
	  {Scaramella}, {Scodeggio}, {Zanichelli}, {Adami}, {Arnouts}, {Bardelli},
	  {Bolzonella}, {Cappi}, {Charlot}, {Contini}, {Foucaud}, {Franzetti},
	  {Garilli}, {Gavignaud}, {Guzzo}, {Ilbert}, {Iovino}, {McCracken}, {Mancini},
	  {Marano}, {Marinoni}, {Mathez}, {Mazure}, {Meneux}, {Merighi}, {Pell{\`o}},
	  {Pollo}, {Pozzetti}, {Radovich}, {Zucca}, {Arnaboldi}, {Bondi}, {Bongiorno},
	  {Busarello}, {Ciliegi}, {Gregorini}, {Mellier}, {Merluzzi}, {Ripepi}, \&
	  {Rizzo}}]{LeFevre2004}
	{Le F{\`e}vre}, O., {Vettolani}, G., {Paltani}, S., {et~al.} 2004, \aap, 428,
	  1043, \dodoi{10.1051/0004-6361:20048072}
	
	\bibitem[{{Le F{\`e}vre} {et~al.}(2005){Le F{\`e}vre}, {Vettolani}, {Garilli},
	  {Tresse}, {Bottini}, {Le Brun}, {Maccagni}, {Picat}, {Scaramella},
	  {Scodeggio}, {Zanichelli}, {Adami}, {Arnaboldi}, {Arnouts}, {Bardelli},
	  {Bolzonella}, {Cappi}, {Charlot}, {Ciliegi}, {Contini}, {Foucaud},
	  {Franzetti}, {Gavignaud}, {Guzzo}, {Ilbert}, {Iovino}, {McCracken}, {Marano},
	  {Marinoni}, {Mathez}, {Mazure}, {Meneux}, {Merighi}, {Paltani}, {Pell{\`o}},
	  {Pollo}, {Pozzetti}, {Radovich}, {Zamorani}, {Zucca}, {Bondi}, {Bongiorno},
	  {Busarello}, {Lamareille}, {Mellier}, {Merluzzi}, {Ripepi}, \&
	  {Rizzo}}]{LeFevre2005}
	{Le F{\`e}vre}, O., {Vettolani}, G., {Garilli}, B., {et~al.} 2005, \aap, 439,
	  845, \dodoi{10.1051/0004-6361:20041960}
	
	\bibitem[{{Le F{\`e}vre} {et~al.}(2013){Le F{\`e}vre}, {Cassata}, {Cucciati},
	  {Garilli}, {Ilbert}, {Le Brun}, {Maccagni}, {Moreau}, {Scodeggio}, {Tresse},
	  {Zamorani}, {Adami}, {Arnouts}, {Bardelli}, {Bolzonella}, {Bondi},
	  {Bongiorno}, {Bottini}, {Cappi}, {Charlot}, {Ciliegi}, {Contini}, {de la
	  Torre}, {Foucaud}, {Franzetti}, {Gavignaud}, {Guzzo}, {Iovino}, {Lemaux},
	  {L{\'o}pez-Sanjuan}, {McCracken}, {Marano}, {Marinoni}, {Mazure}, {Mellier},
	  {Merighi}, {Merluzzi}, {Paltani}, {Pell{\`o}}, {Pollo}, {Pozzetti},
	  {Scaramella}, {Tasca}, {Vergani}, {Vettolani}, {Zanichelli}, \&
	  {Zucca}}]{LeFevre2013}
	{Le F{\`e}vre}, O., {Cassata}, P., {Cucciati}, O., {et~al.} 2013, \aap, 559,
	  A14, \dodoi{10.1051/0004-6361/201322179}
	
	\bibitem[{{Le F{\`e}vre} {et~al.}(2015){Le F{\`e}vre}, {Tasca}, {Cassata},
	  {Garilli}, {Le Brun}, {Maccagni}, {Pentericci}, {Thomas}, {Vanzella},
	  {Zamorani}, {Zucca}, {Amorin}, {Bardelli}, {Capak}, {Cassar{\`a}},
	  {Castellano}, {Cimatti}, {Cuby}, {Cucciati}, {de la Torre}, {Durkalec},
	  {Fontana}, {Giavalisco}, {Grazian}, {Hathi}, {Ilbert}, {Lemaux}, {Moreau},
	  {Paltani}, {Ribeiro}, {Salvato}, {Schaerer}, {Scodeggio}, {Sommariva},
	  {Talia}, {Taniguchi}, {Tresse}, {Vergani}, {Wang}, {Charlot}, {Contini},
	  {Fotopoulou}, {L{\'o}pez-Sanjuan}, {Mellier}, \& {Scoville}}]{LeFevre2015}
	{Le F{\`e}vre}, O., {Tasca}, L.~A.~M., {Cassata}, P., {et~al.} 2015, \aap, 576,
	  A79, \dodoi{10.1051/0004-6361/201423829}
	
	\bibitem[{{Lemaux} {et~al.}(2010){Lemaux}, {Lubin}, {Shapley}, {Kocevski},
	  {Gal}, \& {Squires}}]{Lemaux2010}
	{Lemaux}, B.~C., {Lubin}, L.~M., {Shapley}, A., {et~al.} 2010, \apj, 716, 970,
	  \dodoi{10.1088/0004-637X/716/2/970}
	
	\bibitem[{{Lemaux} {et~al.}(2017){Lemaux}, {Tomczak}, {Lubin}, {Wu}, {Gal},
	  {Rumbaugh}, {Kocevski}, \& {Squires}}]{Lemaux2017}
	{Lemaux}, B.~C., {Tomczak}, A.~R., {Lubin}, L.~M., {et~al.} 2017, \mnras, 472,
	  419, \dodoi{10.1093/mnras/stx1579}
	
	\bibitem[{{Lemaux} {et~al.}(2012){Lemaux}, {Gal}, {Lubin}, {Kocevski},
	  {Fassnacht}, {McGrath}, {Squires}, {Surace}, \& {Lacy}}]{Lemaux2012}
	{Lemaux}, B.~C., {Gal}, R.~R., {Lubin}, L.~M., {et~al.} 2012, \apj, 745, 106,
	  \dodoi{10.1088/0004-637X/745/2/106}
	
	\bibitem[{{Lemaux} {et~al.}(2014{\natexlab{a}}){Lemaux}, {Cucciati}, {Tasca},
	  {Le F{\`e}vre}, {Zamorani}, {Cassata}, {Garilli}, {Le Brun}, {Maccagni},
	  {Pentericci}, {Thomas}, {Vanzella}, {Zucca}, {Amor{\'{\i}}n}, {Bardelli},
	  {Capak}, {Cassar{\`a}}, {Castellano}, {Cimatti}, {Cuby}, {de la Torre},
	  {Durkalec}, {Fontana}, {Giavalisco}, {Grazian}, {Hathi}, {Ilbert}, {Moreau},
	  {Paltani}, {Ribeiro}, {Salvato}, {Schaerer}, {Scodeggio}, {Sommariva},
	  {Talia}, {Taniguchi}, {Tresse}, {Vergani}, {Wang}, {Charlot}, {Contini},
	  {Fotopoulou}, {Gal}, {Kocevski}, {L{\'o}pez-Sanjuan}, {Lubin}, {Mellier},
	  {Sadibekova}, \& {Scoville}}]{Lemaux2014a}
	{Lemaux}, B.~C., {Cucciati}, O., {Tasca}, L.~A.~M., {et~al.}
	  2014{\natexlab{a}}, \aap, 572, A41, \dodoi{10.1051/0004-6361/201423828}
	
	\bibitem[{{Lemaux} {et~al.}(2014{\natexlab{b}}){Lemaux}, {Le Floc'h}, {Le
	  F{\`e}vre}, {Ilbert}, {Tresse}, {Lubin}, {Zamorani}, {Gal}, {Ciliegi},
	  {Cassata}, {Kocevski}, {McGrath}, {Bardelli}, {Zucca}, \&
	  {Squires}}]{Lemaux2014b}
	{Lemaux}, B.~C., {Le Floc'h}, E., {Le F{\`e}vre}, O., {et~al.}
	  2014{\natexlab{b}}, \aap, 572, A90, \dodoi{10.1051/0004-6361/201323089}
	
	\bibitem[{{Lemaux} {et~al.}(2018){Lemaux}, {Le F{\`e}vre}, {Cucciati},
	  {Ribeiro}, {Tasca}, {Zamorani}, {Ilbert}, {Thomas}, {Bardelli}, \&
	  {Cassata}}]{Lemaux2018}
	{Lemaux}, B.~C., {Le F{\`e}vre}, O., {Cucciati}, O., {et~al.} 2018, \aap, 615,
	  A77, \dodoi{10.1051/0004-6361/201730870}
	
	\bibitem[{{Lemaux} {et~al.}(2019){Lemaux}, {Tomczak}, {Lubin}, {Gal}, {Shen},
	  {Pelliccia}, {Wu}, {Hung}, {Mei}, {Le F{\`e}vre}, {Rumbaugh}, {Kocevski}, \&
	  {Squires}}]{Lemaux2019}
	{Lemaux}, B.~C., {Tomczak}, A.~R., {Lubin}, L.~M., {et~al.} 2019, \mnras, 490,
	  1231, \dodoi{10.1093/mnras/stz2661}
	
	\bibitem[{{Lemaux} {et~al.}(2020){Lemaux}, {Cucciati}, {Le F{\`e}vre},
	  {Zamorani}, {Lubin}, {Hathi}, {Ilbert}, {Pelliccia}, {Amor{\'\i}n},
	  {Bardelli}, {Cassata}, {Gal}, {Garilli}, {Guaita}, {Giavalisco}, {Hung},
	  {Koekemoer}, {Maccagni}, {Pentericci}, {Ribeiro}, {Schaerer}, {Shen},
	  {Talia}, {Tomczak}, {Vanzella}, {Vergani}, \& {Zucca}}]{Lemaux2020}
	{Lemaux}, B.~C., {Cucciati}, O., {Le F{\`e}vre}, O., {et~al.} 2020, arXiv
	  e-prints, arXiv:2009.03324.
	\newblock \doarXiv{2009.03324}
	
	\bibitem[{{Lidman} {et~al.}(2012){Lidman}, {Suherli}, {Muzzin}, {Wilson},
	  {Demarco}, {Brough}, {Rettura}, {Cox}, {DeGroot}, {Yee}, {Gilbank},
	  {Hoekstra}, {Balogh}, {Ellingson}, {Hicks}, {Nantais}, {Noble}, {Lacy},
	  {Surace}, \& {Webb}}]{Lidman2012}
	{Lidman}, C., {Suherli}, J., {Muzzin}, A., {et~al.} 2012, \mnras, 427, 550,
	  \dodoi{10.1111/j.1365-2966.2012.21984.x}
	
	\bibitem[{{Lidman} {et~al.}(2013){Lidman}, {Iacobuta}, {Bauer}, {Barrientos},
	  {Cerulo}, {Couch}, {Delaye}, {Demarco}, {Ellingson}, {Faloon}, {Gilbank},
	  {Huertas-Company}, {Mei}, {Meyers}, {Muzzin}, {Noble}, {Nantais}, {Rettura},
	  {Rosati}, {S{\'a}nchez-Janssen}, {Strazzullo}, {Webb}, {Wilson}, {Yan}, \&
	  {Yee}}]{Lidman2013}
	{Lidman}, C., {Iacobuta}, G., {Bauer}, A.~E., {et~al.} 2013, \mnras, 433, 825,
	  \dodoi{10.1093/mnras/stt777}
	
	\bibitem[{{Long} {et~al.}(2020){Long}, {Cooray}, {Ma}, {Casey}, {Wardlow},
	  {Nayyeri}, {Ivison}, {Farrah}, \& {Dannerbauer}}]{Long2020}
	{Long}, A.~S., {Cooray}, A., {Ma}, J., {et~al.} 2020, \apj, 898, 133,
	  \dodoi{10.3847/1538-4357/ab9d1f}
	
	\bibitem[{{Lonsdale} {et~al.}(2003){Lonsdale}, {Smith}, {Rowan-Robinson},
	  {Surace}, {Shupe}, {Xu}, {Oliver}, {Padgett}, {Fang}, {Conrow},
	  {Franceschini}, {Gautier}, {Griffin}, {Hacking}, {Masci}, {Morrison},
	  {O'Linger}, {Owen}, {P{\'e}rez-Fournon}, {Pierre}, {Puetter}, {Stacey},
	  {Castro}, {Polletta}, {Farrah}, {Jarrett}, {Frayer}, {Siana}, {Babbedge},
	  {Dye}, {Fox}, {Gonzalez-Solares}, {Salaman}, {Berta}, {Condon}, {Dole}, \&
	  {Serjeant}}]{Lonsdale2003}
	{Lonsdale}, C.~J., {Smith}, H.~E., {Rowan-Robinson}, M., {et~al.} 2003, \pasp,
	  115, 897, \dodoi{10.1086/376850}
	
	\bibitem[{{Lubin} {et~al.}(2009){Lubin}, {Gal}, {Lemaux}, {Kocevski}, \&
	  {Squires}}]{Lubin2009}
	{Lubin}, L.~M., {Gal}, R.~R., {Lemaux}, B.~C., {Kocevski}, D.~D., \& {Squires},
	  G.~K. 2009, \aj, 137, 4867, \dodoi{10.1088/0004-6256/137/6/4867}
	
	\bibitem[{{Magliocchetti} {et~al.}(2018){Magliocchetti}, {Popesso}, {Brusa}, \&
	  {Salvato}}]{Magliocchetti2018b}
	{Magliocchetti}, M., {Popesso}, P., {Brusa}, M., \& {Salvato}, M. 2018, \mnras,
	  478, 3848, \dodoi{10.1093/mnras/sty1309}
	
	\bibitem[{{Malavasi} {et~al.}(2015){Malavasi}, {Bardelli}, {Ciliegi}, {Ilbert},
	  {Pozzetti}, \& {Zucca}}]{Malavasi2015}
	{Malavasi}, N., {Bardelli}, S., {Ciliegi}, P., {et~al.} 2015, \aap, 576, A101,
	  \dodoi{10.1051/0004-6361/201425155}
	
	\bibitem[{{Matsuda} {et~al.}(2009){Matsuda}, {Nakamura}, {Morimoto}, {Smail},
	  {De Breuck}, {Ohta}, {Kodama}, {Inoue}, {Hayashino}, {Kousai}, {Nakamura},
	  {Horie}, {Yamada}, {Kitamura}, {Saito}, {Taniguchi}, {Tanaka}, \&
	  {Hibon}}]{Matsuda2009}
	{Matsuda}, Y., {Nakamura}, Y., {Morimoto}, N., {et~al.} 2009, \mnras, 400, L66,
	  \dodoi{10.1111/j.1745-3933.2009.00764.x}
	
	\bibitem[{{Mauch} \& {Sadler}(2007)}]{Mauch2007}
	{Mauch}, T., \& {Sadler}, E.~M. 2007, \mnras, 375, 931,
	  \dodoi{10.1111/j.1365-2966.2006.11353.x}
	
	\bibitem[{{McLean} {et~al.}(2012){McLean}, {Steidel}, {Epps}, {Konidaris},
	  {Matthews}, {Adkins}, {Aliado}, {Brims}, {Canfield}, {Cromer}, {Fucik},
	  {Kulas}, {Mace}, {Magnone}, {Rodriguez}, {Rudie}, {Trainor}, {Wang}, {Weber},
	  \& {Weiss}}]{McLean2012}
	{McLean}, I.~S., {Steidel}, C.~C., {Epps}, H.~W., {et~al.} 2012, in Society of
	  Photo-Optical Instrumentation Engineers (SPIE) Conference Series, Vol. 8446,
	  Ground-based and Airborne Instrumentation for Astronomy IV, ed. I.~S.
	  {McLean}, S.~K. {Ramsay}, \& H.~{Takami}, 84460J, \dodoi{10.1117/12.924794}
	
	\bibitem[{{Miley} {et~al.}(2004){Miley}, {Overzier}, {Tsvetanov}, {Bouwens},
	  {Ben{\'\i}tez}, {Blakeslee}, {Ford}, {Illingworth}, {Postman}, {Rosati},
	  {Clampin}, {Hartig}, {Zirm}, {R{\"o}ttgering}, {Venemans}, {Ardila},
	  {Bartko}, {Broadhurst}, {Brown}, {Burrows}, {Cheng}, {Cross}, {De Breuck},
	  {Feldman}, {Franx}, {Golimowski}, {Gronwall}, {Infante}, {Martel},
	  {Menanteau}, {Meurer}, {Sirianni}, {Kimble}, {Krist}, {Sparks}, {Tran},
	  {White}, \& {Zheng}}]{Miley2004}
	{Miley}, G.~K., {Overzier}, R.~A., {Tsvetanov}, Z.~I., {et~al.} 2004, \nat,
	  427, 47, \dodoi{10.1038/nature02125}
	
	\bibitem[{{Miller} \& {Owen}(2002)}]{Miller2002}
	{Miller}, N.~A., \& {Owen}, F.~N. 2002, \aj, 124, 2453, \dodoi{10.1086/343837}
	
	\bibitem[{{Muzzin} {et~al.}(2013){Muzzin}, {Marchesini}, {Stefanon}, {Franx},
	  {McCracken}, {Milvang-Jensen}, {Dunlop}, {Fynbo}, {Brammer}, {Labb{\'e}}, \&
	  {van Dokkum}}]{Muzzin2013}
	{Muzzin}, A., {Marchesini}, D., {Stefanon}, M., {et~al.} 2013, \apj, 777, 18,
	  \dodoi{10.1088/0004-637X/777/1/18}
	
	\bibitem[{{Newman} {et~al.}(2013){Newman}, {Cooper}, {Davis}, {Faber}, {Coil},
	  {Guhathakurta}, {Koo}, {Phillips}, {Conroy}, {Dutton}, {Finkbeiner}, {Gerke},
	  {Rosario}, {Weiner}, {Willmer}, {Yan}, {Harker}, {Kassin}, {Konidaris},
	  {Lai}, {Madgwick}, {Noeske}, {Wirth}, {Connolly}, {Kaiser}, {Kirby},
	  {Lemaux}, {Lin}, {Lotz}, {Luppino}, {Marinoni}, {Matthews}, {Metevier}, \&
	  {Schiavon}}]{Newman2013}
	{Newman}, J.~A., {Cooper}, M.~C., {Davis}, M., {et~al.} 2013, \apjs, 208, 5,
	  \dodoi{10.1088/0067-0049/208/1/5}
	
	\bibitem[{{Oke} \& {Gunn}(1983)}]{Oke1983}
	{Oke}, J.~B., \& {Gunn}, J.~E. 1983, \apj, 266, 713, \dodoi{10.1086/160817}
	
	\bibitem[{{Old} {et~al.}(2020){Old}, {Balogh}, {van der Burg}, {Biviano},
	  {Yee}, {Pintos-Castro}, {Webb}, {Muzzin}, {Rudnick}, {Vulcani}, {Poggianti},
	  {Cooper}, {Zaritsky}, {Cerulo}, {Wilson}, {Chan}, {Lidman}, {McGee},
	  {Demarco}, {Forrest}, {De Lucia}, {Gilbank}, {Kukstas}, {McCarthy},
	  {Jablonka}, {Nantais}, {Noble}, {Reeves}, \& {Shipley}}]{Old2020}
	{Old}, L.~J., {Balogh}, M.~L., {van der Burg}, R. F.~J., {et~al.} 2020, \mnras,
	  493, 5987, \dodoi{10.1093/mnras/staa579}
	
	\bibitem[{{Orsi} {et~al.}(2016){Orsi}, {Fanidakis}, {Lacey}, \&
	  {Baugh}}]{Orsi2016}
	{Orsi}, {\'A}.~A., {Fanidakis}, N., {Lacey}, C.~G., \& {Baugh}, C.~M. 2016,
	  \mnras, 456, 3827, \dodoi{10.1093/mnras/stv2919}
	
	\bibitem[{{Overzier}(2016)}]{Overzier2016}
	{Overzier}, R.~A. 2016, \aapr, 24, 14, \dodoi{10.1007/s00159-016-0100-3}
	
	\bibitem[{{Overzier} {et~al.}(2008){Overzier}, {Bouwens}, {Cross}, {Venemans},
	  {Miley}, {Zirm}, {Ben{\'\i}tez}, {Blakeslee}, {Coe}, {Demarco}, {Ford},
	  {Homeier}, {Illingworth}, {Kurk}, {Martel}, {Mei}, {Oliveira},
	  {R{\"o}ttgering}, {Tsvetanov}, \& {Zheng}}]{Overzier2008}
	{Overzier}, R.~A., {Bouwens}, R.~J., {Cross}, N.~J.~G., {et~al.} 2008, \apj,
	  673, 143, \dodoi{10.1086/524342}
	
	\bibitem[{{Overzier} {et~al.}(2009){Overzier}, {Shu}, {Zheng}, {Rettura},
	  {Zirm}, {Bouwens}, {Ford}, {Illingworth}, {Miley}, {Venemans}, \&
	  {White}}]{Overzier2009}
	{Overzier}, R.~A., {Shu}, X., {Zheng}, W., {et~al.} 2009, \apj, 704, 548,
	  \dodoi{10.1088/0004-637X/704/1/548}
	
	\bibitem[{{Padmanabhan}(1993)}]{Padmanabhan1993}
	{Padmanabhan}, T. 1993, {Structure Formation in the Universe}
	
	\bibitem[{{Padovani}(2016)}]{Padovani2016}
	{Padovani}, P. 2016, \aapr, 24, 13, \dodoi{10.1007/s00159-016-0098-6}
	
	\bibitem[{{Peng} {et~al.}(2010){Peng}, {Lilly}, {Kova{\v c}}, {Bolzonella},
	  {Pozzetti}, {Renzini}, {Zamorani}, {Ilbert}, {Knobel}, {Iovino}, {Maier},
	  {Cucciati}, {Tasca}, {Carollo}, {Silverman}, {Kampczyk}, {de Ravel},
	  {Sanders}, {Scoville}, {Contini}, {Mainieri}, {Scodeggio}, {Kneib}, {Le
	  F{\`e}vre}, {Bardelli}, {Bongiorno}, {Caputi}, {Coppa}, {de la Torre},
	  {Franzetti}, {Garilli}, {Lamareille}, {Le Borgne}, {Le Brun}, {Mignoli},
	  {Perez Montero}, {Pello}, {Ricciardelli}, {Tanaka}, {Tresse}, {Vergani},
	  {Welikala}, {Zucca}, {Oesch}, {Abbas}, {Barnes}, {Bordoloi}, {Bottini},
	  {Cappi}, {Cassata}, {Cimatti}, {Fumana}, {Hasinger}, {Koekemoer},
	  {Leauthaud}, {Maccagni}, {Marinoni}, {McCracken}, {Memeo}, {Meneux}, {Nair},
	  {Porciani}, {Presotto}, \& {Scaramella}}]{Peng2010}
	{Peng}, Y.-j., {Lilly}, S.~J., {Kova{\v c}}, K., {et~al.} 2010, \apj, 721, 193,
	  \dodoi{10.1088/0004-637X/721/1/193}
	
	\bibitem[{{Pentericci} {et~al.}(2000){Pentericci}, {Kurk}, {R{\"o}ttgering},
	  {Miley}, {van Breugel}, {Carilli}, {Ford}, {Heckman}, {McCarthy}, \&
	  {Moorwood}}]{Pentericci2000}
	{Pentericci}, L., {Kurk}, J.~D., {R{\"o}ttgering}, H.~J.~A., {et~al.} 2000,
	  \aap, 361, L25.
	\newblock \doarXiv{astro-ph/0008143}
	
	\bibitem[{{Pilbratt} {et~al.}(2010){Pilbratt}, {Riedinger}, {Passvogel},
	  {Crone}, {Doyle}, {Gageur}, {Heras}, {Jewell}, {Metcalfe}, {Ott}, \&
	  {Schmidt}}]{Pilbratt2010}
	{Pilbratt}, G.~L., {Riedinger}, J.~R., {Passvogel}, T., {et~al.} 2010, \aap,
	  518, L1, \dodoi{10.1051/0004-6361/201014759}
	
	\bibitem[{{Puget} {et~al.}(2004){Puget}, {Stadler}, {Doyon}, {Gigan},
	  {Thibault}, {Luppino}, {Barrick}, {Benedict}, {Forveille}, {Rambold},
	  {Thomas}, {Vermeulen}, {Ward}, {Beuzit}, {Feautrier}, {Magnard}, {Mella},
	  {Preis}, {Vallee}, {Wang}, {Lin}, {Hall}, \& {Hodapp}}]{Puget2004}
	{Puget}, P., {Stadler}, E., {Doyon}, R., {et~al.} 2004, in \procspie, Vol.
	  5492, Ground-based Instrumentation for Astronomy, ed. A.~F.~M. {Moorwood} \&
	  M.~{Iye}, 978--987, \dodoi{10.1117/12.551097}
	
	\bibitem[{{Rennehan} {et~al.}(2020){Rennehan}, {Babul}, {Hayward}, {Bottrell},
	  {Hani}, \& {Chapman}}]{Rennehan2020}
	{Rennehan}, D., {Babul}, A., {Hayward}, C.~C., {et~al.} 2020, \mnras, 493,
	  4607, \dodoi{10.1093/mnras/staa541}
	
	\bibitem[{{Rieke} {et~al.}(2004){Rieke}, {Young}, {Engelbracht}, {Kelly},
	  {Low}, {Haller}, {Beeman}, {Gordon}, {Stansberry}, {Misselt}, {Cadien},
	  {Morrison}, {Rivlis}, {Latter}, {Noriega-Crespo}, {Padgett}, {Stapelfeldt},
	  {Hines}, {Egami}, {Muzerolle}, {Alonso-Herrero}, {Blaylock}, {Dole}, {Hinz},
	  {Le Floc'h}, {Papovich}, {P{\'e}rez-Gonz{\'a}lez}, {Smith}, {Su}, {Bennett},
	  {Frayer}, {Henderson}, {Lu}, {Masci}, {Pesenson}, {Rebull}, {Rho}, {Keene},
	  {Stolovy}, {Wachter}, {Wheaton}, {Werner}, \& {Richards}}]{Rieke2004}
	{Rieke}, G.~H., {Young}, E.~T., {Engelbracht}, C.~W., {et~al.} 2004, \apjs,
	  154, 25, \dodoi{10.1086/422717}
	
	\bibitem[{{Rigby} {et~al.}(2014){Rigby}, {Hatch}, {R{\"o}ttgering},
	  {Sibthorpe}, {Chiang}, {Overzier}, {Herbonnet}, {Borgani}, {Clements},
	  {Dannerbauer}, {De Breuck}, {De Lucia}, {Kurk}, {Maschietto}, {Miley},
	  {Saro}, {Seymour}, \& {Venemans}}]{Rigby2014}
	{Rigby}, E.~E., {Hatch}, N.~A., {R{\"o}ttgering}, H.~J.~A., {et~al.} 2014,
	  \mnras, 437, 1882, \dodoi{10.1093/mnras/stt2019}
	
	\bibitem[{{R{\"o}ttgering} {et~al.}(2003){R{\"o}ttgering}, {Daddi}, {Overzier},
	  \& {Wilman}}]{Rottgering2003}
	{R{\"o}ttgering}, H., {Daddi}, E., {Overzier}, R., \& {Wilman}, R. 2003, \nar,
	  47, 309, \dodoi{10.1016/S1387-6473(03)00129-5}
	
	\bibitem[{{Rumbaugh} {et~al.}(2012){Rumbaugh}, {Kocevski}, {Gal}, {Lemaux},
	  {Lubin}, {Fassnacht}, {McGrath}, \& {Squires}}]{Rumbaugh2012}
	{Rumbaugh}, N., {Kocevski}, D.~D., {Gal}, R.~R., {et~al.} 2012, \apj, 746, 155,
	  \dodoi{10.1088/0004-637X/746/2/155}
	
	\bibitem[{{Shah} {et~al.}(2020){Shah}, {Kartaltepe}, {Magagnoli}, {Cox},
	  {Wetherell}, {Vanderhoof}, {Calabro}, {Chartab}, {Conselice}, {Croton},
	  {Donley}, {de Groot}, {de la Vega}, {Hathi}, {Ilbert}, {Inami}, {Kocevski},
	  {Koekemoer}, {Lemaux}, {Mantha}, {Marchesi}, {Martig}, {Masters}, {McGrath},
	  {McIntosh}, {Moreno}, {Nayyeri}, {Pampliega}, {Salvato}, {Snyder},
	  {Straughn}, {Treister}, \& {Weston}}]{Shah2020}
	{Shah}, E.~A., {Kartaltepe}, J.~S., {Magagnoli}, C.~T., {et~al.} 2020, \apj,
	  904, 107, \dodoi{10.3847/1538-4357/abbf59}
	
	\bibitem[{{Shen} {et~al.}(2017){Shen}, {Miller}, {Lemaux}, {Tomczak}, {Lubin},
	  {Rumbaugh}, {Fassnacht}, {Becker}, {Gal}, {Wu}, \& {Squires}}]{Shen2017}
	{Shen}, L., {Miller}, N.~A., {Lemaux}, B.~C., {et~al.} 2017, \mnras, 472, 998,
	  \dodoi{10.1093/mnras/stx1984}
	
	\bibitem[{{Shen} {et~al.}(2019){Shen}, {Tomczak}, {Lemaux}, {Pelliccia},
	  {Lubin}, {Miller}, {Perrotta}, {Fassnacht}, {Becker}, \& {Gal}}]{Shen2019}
	{Shen}, L., {Tomczak}, A.~R., {Lemaux}, B.~C., {et~al.} 2019, \mnras, 484,
	  2433, \dodoi{10.1093/mnras/stz152}
	
	\bibitem[{{Shen} {et~al.}(2020{\natexlab{a}}){Shen}, {Lemaux}, {Lubin},
	  {McKean}, {Miller}, {Pelliccia}, {Fassnacht}, {Tomczak}, {Wu}, {Kocevski},
	  {Gal}, {Hung}, \& {Squires}}]{Shen2020a}
	{Shen}, L., {Lemaux}, B.~C., {Lubin}, L.~M., {et~al.} 2020{\natexlab{a}},
	  \mnras, 494, 5374, \dodoi{10.1093/mnras/staa1005}
	
	\bibitem[{{Shen} {et~al.}(2020{\natexlab{b}}){Shen}, {Liu}, {Zhang}, {Lemaux},
	  {Lubin}, {Pelliccia}, {Moravec}, {Golden-Marx}, {Zhou}, {Fang}, {Tomczak},
	  {McKean}, {Miller}, {Fassnacht}, {Wu}, {Kocevski}, {Gal}, {Hung}, \&
	  {Squires}}]{Shen2020b}
	{Shen}, L., {Liu}, G., {Zhang}, M., {et~al.} 2020{\natexlab{b}}, arXiv
	  e-prints, arXiv:2009.03343.
	\newblock \doarXiv{2009.03343}
	
	\bibitem[{{Shimakawa} {et~al.}(2014){Shimakawa}, {Kodama}, {Tadaki}, {Tanaka},
	  {Hayashi}, \& {Koyama}}]{Shimakawa2014}
	{Shimakawa}, R., {Kodama}, T., {Tadaki}, K.~I., {et~al.} 2014, \mnras, 441, L1,
	  \dodoi{10.1093/mnrasl/slu029}
	
	\bibitem[{{Smol{\v{c}}i{\'c}} {et~al.}(2009){Smol{\v{c}}i{\'c}}, {Schinnerer},
	  {Zamorani}, {Bell}, {Bondi}, {Carilli}, {Ciliegi}, {Mobasher}, {Paglione},
	  {Scodeggio}, \& {Scoville}}]{Smolcic2009a}
	{Smol{\v{c}}i{\'c}}, V., {Schinnerer}, E., {Zamorani}, G., {et~al.} 2009, \apj,
	  690, 610, \dodoi{10.1088/0004-637X/690/1/610}
	
	\bibitem[{{Steidel} {et~al.}(1998){Steidel}, {Adelberger}, {Dickinson},
	  {Giavalisco}, {Pettini}, \& {Kellogg}}]{Steidel1998}
	{Steidel}, C.~C., {Adelberger}, K.~L., {Dickinson}, M., {et~al.} 1998, \apj,
	  492, 428, \dodoi{10.1086/305073}
	
	\bibitem[{{Tadhunter}(2016)}]{Tadhunter2016}
	{Tadhunter}, C. 2016, \aapr, 24, 10, \dodoi{10.1007/s00159-016-0094-x}
	
	\bibitem[{{Thomas} {et~al.}(2017){Thomas}, {Le F{\`e}vre}, {Scodeggio},
	  {Cassata}, {Garilli}, {Le Brun}, {Lemaux}, {Maccagni}, {Pforr}, {Tasca},
	  {Zamorani}, {Bardelli}, {Hathi}, {Tresse}, {Zucca}, \&
	  {Koekemoer}}]{Thomas2017}
	{Thomas}, R., {Le F{\`e}vre}, O., {Scodeggio}, M., {et~al.} 2017, \aap, 602,
	  A35, \dodoi{10.1051/0004-6361/201628141}
	
	\bibitem[{{Tomczak} {et~al.}(2017){Tomczak}, {Lemaux}, {Lubin}, {Gal}, {Wu},
	  {Holden}, {Kocevski}, {Mei}, {Pelliccia}, {Rumbaugh}, \&
	  {Shen}}]{Tomczak2017}
	{Tomczak}, A.~R., {Lemaux}, B.~C., {Lubin}, L.~M., {et~al.} 2017, ArXiv
	  e-prints.
	\newblock \doarXiv{1709.00011}
	
	\bibitem[{{Tomczak} {et~al.}(2019){Tomczak}, {Lemaux}, {Lubin}, {Pelliccia},
	  {Shen}, {Gal}, {Hung}, {Kocevski}, {Le F{\`e}vre}, {Mei}, {Rumbaugh},
	  {Squires}, \& {Wu}}]{Tomczak2019}
	---. 2019, \mnras, 484, 4695, \dodoi{10.1093/mnras/stz342}
	
	\bibitem[{{Toshikawa} {et~al.}(2020){Toshikawa}, {Malkan}, {Kashikawa},
	  {Overzier}, {Uchiyama}, {Ota}, {Ishikawa}, \& {Ito}}]{Toshikawa2020}
	{Toshikawa}, J., {Malkan}, M.~A., {Kashikawa}, N., {et~al.} 2020, \apj, 888,
	  89, \dodoi{10.3847/1538-4357/ab5e85}
	
	\bibitem[{{Toshikawa} {et~al.}(2016){Toshikawa}, {Kashikawa}, {Overzier},
	  {Malkan}, {Furusawa}, {Ishikawa}, {Onoue}, {Ota}, {Tanaka}, {Niino}, \&
	  {Uchiyama}}]{Toshikawa2016}
	{Toshikawa}, J., {Kashikawa}, N., {Overzier}, R., {et~al.} 2016, \apj, 826,
	  114, \dodoi{10.3847/0004-637X/826/2/114}
	
	\bibitem[{{Toshikawa} {et~al.}(2018){Toshikawa}, {Uchiyama}, {Kashikawa},
	  {Ouchi}, {Overzier}, {Ono}, {Harikane}, {Ishikawa}, {Kodama}, {Matsuda},
	  {Lin}, {Onoue}, {Tanaka}, {Nagao}, {Akiyama}, {Komiyama}, {Goto}, \&
	  {Lee}}]{Toshikawa2018}
	{Toshikawa}, J., {Uchiyama}, H., {Kashikawa}, N., {et~al.} 2018, \pasj, 70,
	  S12, \dodoi{10.1093/pasj/psx102}
	
	\bibitem[{{Valentino} {et~al.}(2016){Valentino}, {Daddi}, {Finoguenov},
	  {Strazzullo}, {Le Brun}, {Vignali}, {Bournaud}, {Dickinson}, {Renzini},
	  {B{\'e}thermin}, {Zanella}, {Gobat}, {Cimatti}, {Elbaz}, {Onodera},
	  {Pannella}, {Sargent}, {Arimoto}, {Carollo}, \& {Starck}}]{Valentino2016}
	{Valentino}, F., {Daddi}, E., {Finoguenov}, A., {et~al.} 2016, \apj, 829, 53,
	  \dodoi{10.3847/0004-637X/829/1/53}
	
	\bibitem[{{Valtchanov} {et~al.}(2004){Valtchanov}, {Pierre}, {Willis}, {Dos
	  Santos}, {Jones}, {Andreon}, {Adami}, {Altieri}, {Bolzonella}, {Bremer},
	  {Duc}, {Gosset}, {Jean}, \& {Surdej}}]{Valtchanov2004}
	{Valtchanov}, I., {Pierre}, M., {Willis}, J., {et~al.} 2004, \aap, 423, 75,
	  \dodoi{10.1051/0004-6361:20040162}
	
	\bibitem[{{van der Burg} {et~al.}(2020){van der Burg}, {Rudnick}, {Balogh},
	  {Muzzin}, {Lidman}, {Old}, {Shipley}, {Gilbank}, {McGee}, {Biviano},
	  {Cerulo}, {Chan}, {Cooper}, {De Lucia}, {Demarco}, {Forrest}, {Gwyn},
	  {Jablonka}, {Kukstas}, {Marchesini}, {Nantais}, {Noble}, {Pintos-Castro},
	  {Poggianti}, {Reeves}, {Stefanon}, {Vulcani}, {Webb}, {Wilson}, {Yee}, \&
	  {Zaritsky}}]{VanderBurg2020}
	{van der Burg}, R. F.~J., {Rudnick}, G., {Balogh}, M.~L., {et~al.} 2020, \aap,
	  638, A112, \dodoi{10.1051/0004-6361/202037754}
	
	\bibitem[{{Venemans} {et~al.}(2002){Venemans}, {Kurk}, {Miley},
	  {R{\"o}ttgering}, {van Breugel}, {Carilli}, {De Breuck}, {Ford}, {Heckman},
	  {McCarthy}, \& {Pentericci}}]{Venemans2002}
	{Venemans}, B.~P., {Kurk}, J.~D., {Miley}, G.~K., {et~al.} 2002, \apjl, 569,
	  L11, \dodoi{10.1086/340563}
	
	\bibitem[{{Venemans} {et~al.}(2007){Venemans}, {R{\"o}ttgering}, {Miley}, {van
	  Breugel}, {de Breuck}, {Kurk}, {Pentericci}, {Stanford}, {Overzier}, {Croft},
	  \& {Ford}}]{Venemans2007}
	{Venemans}, B.~P., {R{\"o}ttgering}, H.~J.~A., {Miley}, G.~K., {et~al.} 2007,
	  \aap, 461, 823, \dodoi{10.1051/0004-6361:20053941}
	
	\bibitem[{{Wang} {et~al.}(2016){Wang}, {Elbaz}, {Daddi}, {Finoguenov}, {Liu},
	  {Schreiber}, {Mart{\'\i}n}, {Strazzullo}, {Valentino}, {van der Burg},
	  {Zanella}, {Ciesla}, {Gobat}, {Le Brun}, {Pannella}, {Sargent}, {Shu}, {Tan},
	  {Cappelluti}, \& {Li}}]{Wang2016}
	{Wang}, T., {Elbaz}, D., {Daddi}, E., {et~al.} 2016, \apj, 828, 56,
	  \dodoi{10.3847/0004-637X/828/1/56}
	
	\bibitem[{{Wylezalek} {et~al.}(2013){Wylezalek}, {Galametz}, {Stern}, {Vernet},
	  {De Breuck}, {Seymour}, {Brodwin}, {Eisenhardt}, {Gonzalez}, {Hatch},
	  {Jarvis}, {Rettura}, {Stanford}, \& {Stevens}}]{Wylezalek2013}
	{Wylezalek}, D., {Galametz}, A., {Stern}, D., {et~al.} 2013, \apj, 769, 79,
	  \dodoi{10.1088/0004-637X/769/1/79}
	
	\bibitem[{{Wylezalek} {et~al.}(2014){Wylezalek}, {Vernet}, {De Breuck},
	  {Stern}, {Brodwin}, {Galametz}, {Gonzalez}, {Jarvis}, {Hatch}, {Seymour}, \&
	  {Stanford}}]{Wylezalek2014}
	{Wylezalek}, D., {Vernet}, J., {De Breuck}, C., {et~al.} 2014, \apj, 786, 17,
	  \dodoi{10.1088/0004-637X/786/1/17}
	
	\bibitem[{{Yan} {et~al.}(2006){Yan}, {Newman}, {Faber}, {Konidaris}, {Koo}, \&
	  {Davis}}]{Yan2006}
	{Yan}, R., {Newman}, J.~A., {Faber}, S.~M., {et~al.} 2006, \apj, 648, 281,
	  \dodoi{10.1086/505629}
	
	\end{thebibliography}



\end{document}